\documentclass[ reprint, nofootinbib, amsmath,amssymb, aps, pra ]{revtex4-2}
\usepackage[a4paper,top=3cm,bottom=3cm,left=2.0cm,right=2.0cm]{geometry}
\usepackage[english]{babel}
\usepackage[T1]{fontenc}
\usepackage{amssymb}
\usepackage{amsfonts,mathrsfs}
\usepackage{amsthm}
\usepackage{hyperref}
\usepackage{graphicx}
\usepackage{dcolumn}
\usepackage{bm}
\usepackage{tabularx}
\usepackage{array}
\usepackage{float}
\usepackage{braket}
\usepackage[utf8]{inputenc}
\usepackage[usenames,dvipsnames]{xcolor}
\usepackage{afterpage}

\renewcommand{\u}{\"{u}}

\theoremstyle{definition}
\newtheorem{definition}{Definition}[section]

\theoremstyle{remark}

\newcommand{\proj}[2]{\Ket{#1}\Bra{#2}}

\begin{document}


\title{Theoretical framework for physical unclonable functions, including quantum readout}

\author{Giulio Gianfelici}
\email{giulio.gianfelici@uni-duesseldorf.de}
\author{Hermann Kampermann}
\author{Dagmar Bru\ss}
\affiliation{ Institut f\"{u}r Theoretische Physik III, Heinrich-Heine-Universit\"{a}t D\"{u}sseldorf, D-40225 D\"{u}sseldorf, Germany }

\date{\today}

\begin{abstract}
\noindent
We propose a theoretical framework to quantitatively describe Physical Unclonable Functions (PUFs), including extensions to quantum protocols, so-called Quantum Readout PUFs (QR-PUFs).
$\text{(QR-)}$ PUFs are physical systems with challenge-response behavior intended to be hard to clone or simulate. Their use has been proposed in several cryptographic protocols, with particular emphasis on authentication.
Here, we provide theoretical assumptions and definitions behind the intuitive ideas of $\text{(QR-)}$ PUFs. This allows to quantitatively characterize the security of such devices in cryptographic protocols. First, by
generalizing previous ideas, we design a general authentication scheme, which is applicable to different physical implementations of both classical PUFs and $\text{(QR-)}$ PUFs. Then, we define the \emph{robustness} and the \emph{unclonability}, which allows us to derive security thresholds for $\text{(QR-)}$ PUF authentication and paves the way to develop further new authentication protocols.
\end{abstract}

\maketitle

\section{Introduction}
\label{sec:Intro}
\emph{Authentication} is a major task of both classical and quantum cryptography. To achieve secure communication between two parties Alice and Bob, it is necessary to ensure that no intruder may participate in the communication, pretending to be one of the legitimate parties, e.g. by a so-called \emph{Man-in-the-middle attack} \cite{KM}. Authentication is ultimately classical, even in quantum protocols like QKD \cite{SBCDLP}.

The main ingredient of an authentication protocol is a shared secret between the legitimate parties: during any authenticated communication Alice and Bob must prove the possession of this secret to confirm their identity.
One has to distinguish two types of authentication \cite{KM}. \emph{Message authentication} is the assurance that a given entity was the original source of the received data. This type of authentication can be achieved by unconditionally secure protocols \cite{WC}. \emph{Entity authentication} is the assurance that a given entity can prove its identity and its involvement in the communication session to another entity. 

Entity authentication is particularly important if there is an asymmetry between the parties, e.g. when one party, namely Alice, is a trusted institution and the other one, namely Bob, is an untrusted user. The communication between Alice and Bob may happen on an authenticated channel owned by Alice, where Bob interacts through a remote terminal. In that case, a one-way entity authentication protocol will be used by Alice to authenticate Bob and to allow him to use her channel.
Such protocols are usually based on a \emph{challenge-response authentication}, a type of authentication where Alice presents a \emph{challenge} and Bob provides a valid \emph{response}, based on the common secret, to be authenticated. For instance, Alice can ask for a password (challenge) and Bob will provide the correct one (response).

In the case of asymmetric communication, it is useful to design authentication protocols based on something the parties possess. 
The trusted Alice can still be required to have secret knowledge since she is able to conceal information from an adversary, but Bob is required only to protect a given token from theft. 
A crucial condition of this approach is that the object has to be unique and an adversary, namely Eve, should not be able to copy it easily.

A \emph{Physical Unclonable Function} (PUF) \cite{RP} is a physical system which can interact in a very complex way with an external signal (which can serve as a challenge) to give an unpredictable output (which can serve as a response). Its internal disorder is exploited to make it unique, hard to clone or simulate.
PUFs are particularly suited for entity authentication because their internal structure plays the role of the shared secret. They can also be used in other protocols, like oblivious transfer \cite{UR10}, bit commitment \cite{RD} or classical key distribution \cite{BFSK}. 
There is a large variety of PUFs, such as the \emph{Optical PUF} \cite{PRTG}, the \emph{Arbiter PUF} \cite{LLGSDD}, the \emph{SRAM PUF} \cite{GKST}, the \emph{Coating PUF} \cite{TSSGVW}, the \emph{Magnetic PUF} \cite{IM}, the \emph{Ring Oscillator PUF} \cite{BNCF} and so on. A more detailed description of the whole family of PUFs is given in \cite{MBWRY} and in \cite{MV}.

To ensure reliability and security it is required to post-process the PUFs' outputs \cite{DGSV, PMBHS}. The most common way to do it is by using the so-called \emph{fuzzy extractor} \cite{DORS}, a tool which combines error correction and privacy amplification.
Error correction is necessary because the PUF's output can be different each time the PUF interacts with the same challenge, even when the authentication involves the real Bob with the original PUF. This can be due to an erroneous implementation of the challenge or to noise in the physical process.
Privacy amplification is important since the outcomes of a PUF are generally non-uniform, i.e. there exist correlations between different responses that can be used by an adversary to undermine the PUF's security. Furthermore, the response, once it is mapped into a uniform key, can, in principle, be used in different protocols other than entity authentication. 

However, even when dealing with noise and non-uniformity, there are some issues with PUFs, because it has been shown that many of them can be actually cloned or simulated \cite{HBNS, RSSDDS, R-etal}, compromising their use in secure authentication schemes.

To solve these problems, an extension of PUFs to quantum protocols was suggested, the so-called \emph{Quantum Readout PUFs} (QR-PUFs) \cite{BS}. Such PUFs encode challenges and responses in quantum states, thus they are expected to be more secure and reliable than classical PUFs, as they add a layer of complexity given by the unclonability of the involved quantum states \cite{WZ}. Moreover, if such quantum states are non-orthogonal, an adversary cannot perfectly distinguish them, and an attempt to do it would introduce disturbances, thus exposing the presence of an intruder to the legitimate parties. 

It is desirable to establish a theoretical framework in which one can perform a rigorous, quantitative, analysis of the security properties of $\text{(QR-)}$ PUFs. Several efforts have been made to formalize the intuitive ideas of PUF \cite{RSS, AMSSW, PK, PM, JD}, and they all capture some aspects of them, but a well-defined agreement about theoretical assumptions and definitions is still lacking. Moreover, the previous approaches are devoted to classical PUFs only.

In this article we propose a common theoretical framework by quantitatively characterizing the $\text{(QR-)}$ PUF properties, particularly the \emph{robustness} \cite{AMSSW} against noise and the \emph{unclonability}. This is done by generalizing ideas from previous approaches (in particular from \cite{AMSSW}) to encompass both classical and QR-PUFs.
Moreover, we introduce a generic scheme for authentication protocols with $\text{(QR-)}$ PUFs, for which security thresholds can be calculated once an experimental implementation is specified. This scheme provides an abstract formalization of existing protocols, together with new ideas such as the difference between a \textit{physical layer} and a \textit{mathematical layer} (see Sec. \ref{sec:Auth}) or the concept of the \textit{shifter} (see Secs. \ref{subsec:cenr} and \ref{subsec:qenr}). 
This framework is designed to be independent of the specific experimental implementation, such that a comparison of different types of PUFs and QR-PUFs becomes possible.
In particular, all implementations use a fuzzy extractor for post-processing.
We expect that this analysis supports both theoretical and experimental research on $\text{(QR-)}$ PUFs, by promoting the implementation of such devices in existing and new secure authentication schemes.

The paper is organized as follows.
In Sec. \ref{sec:Auth} we give an introduction on entity authentication protocols with $\text{(QR-)}$ PUFs. Sec. \ref{sec:not} contains the notation we will use in the paper, in Sec. \ref{sec:class} we describe a protocol with a generic classical PUF, and in Sec. \ref{sec:quant} we generalize this to a generic QR-PUF. The shared formalization of the theoretical properties of $\text{(QR-)}$ PUFs is stated in Sec. \ref{sec:prop} and the formalism is applied in some examples in Sec. \ref{sec:ex}. Some final remarks and the outlook of the work are given in the Conclusion.

\section{Authentication protocols}
\label{sec:Auth}
In the following, we will always call Alice the party that has to authenticate Bob. Mutual authentication can be achieved by repeating the protocol swapping the roles of Alice and Bob. Moreover, we stated in the Introduction that the raw output of a $\text{(QR-)}$ PUF has to be post-processed to be used in secure cryptographic protocols. Therefore, for the sake of clarity, we call \emph{outcome} the raw output while we mean with \emph{response} only the post-processed uniform key.

\begin{figure}[htbp]
\centering
\includegraphics[width=0.48\textwidth]{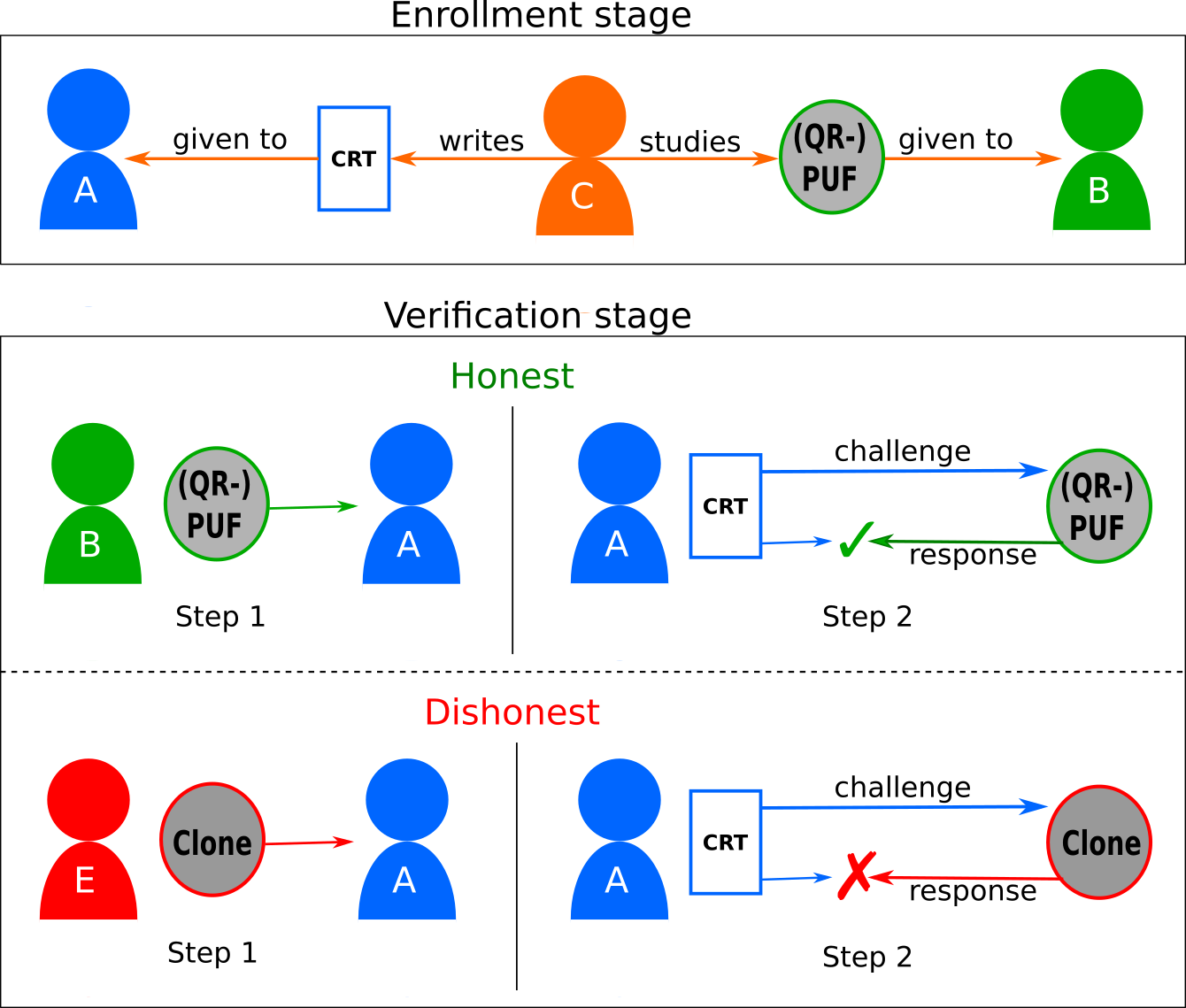}
\caption{ A schematic description of the authentication scheme (colour online). \\
 \textbf{Top:} Enrollment stage. The Certifier (C, orange) studies the $\text{(QR-)}$ PUF's properties and generates the Challenge-Response Table (CRT). Then the CRT is given to Alice (A, blue) and the $\text{(QR-)}$ PUF is given to Bob (B, green).\\
 \textbf{Bottom:} Verification stage. In the honest case, Bob lets Alice interact with his $\text{(QR-)}$ PUF through a terminal and she remotely verifies his identity with the CRT, thus authenticating him. In the dishonest case, an adversary Eve (E, red) claims to be Bob, letting Alice interact with a clone of the $\text{(QR-)}$ PUF, and the protocol should lead to an abortion.}
\label{fig:enver}
\end{figure}

Entity authentication protocols with $\text{(QR-)}$ PUFs consist of two phases \cite{STO}, the \emph{enrollment stage} and the \emph{verification stage} (see fig. \ref{fig:enver}).

The enrollment stage is a part of the protocol which happens only once at the beginning, after the manufacture of the $\text{(QR-)}$ PUF and before any communications between Alice and Bob. An entity, or a group of entities, called the \emph{$\text{(QR-)}$ PUF Certifier} (which may be the $\text{(QR-)}$ PUF manufacturer, Alice itself, a third trusted party or a combination of all of them) studies the $\text{(QR-)}$ PUF's properties, evaluates the parameters needed for the implementation and the post-processing. 
In particular, the Certifier selects a certain number $N$ of challenges and records the corresponding responses.
Challenges and responses form the so-called \emph{Challenge-Response pairs} (CRPs) and they are stored as a \emph{Challenge-Response Table} (CRT), together with additional information needed in the remaining part of the protocol.
After the end of this stage, the Certifier gives the CRT to Alice (which then \emph{knows} the secret) and the $\text{(QR-)}$ PUF to Bob (which then \emph{has} the secret).

The verification stage is the part of the protocol where communication between Alice and Bob is necessary. In this stage, Bob declares his identity to Alice with his $\text{(QR-)}$ PUF, remotely interacting with her through her terminal. To authenticate Bob, Alice sends randomly one challenge from the CRT to the $\text{(QR-)}$ PUF and collects the outcome, which is then post-processed. The calculated response is compared with the one in the CRT, i.e. the one obtained in the enrollment stage. If they match, Alice authenticates Bob.
This stage can be repeated every time Alice needs to authenticate Bob. After every round, however, the used challenge-response pair has to be eliminated from the CRT and cannot be used again \footnote{It was argued \cite{BS} that in the QR-PUF case, challenge-response pairs could be used again, because an adversary is not able to gain full information about their state. Such claims need to be quantitatively proven, here we continue as if any reused CRP is insecure.}.

Depending on the different types of $\text{(QR-)}$ PUFs, the challenges could be different types of physical quantities. For instance, optical PUFs are transparent materials filled with light scattering particles: a laser that interacts with one of them is turned into a unique speckle pattern. For a classical optical PUF, the challenge is the laser orientation and the outcome is the intensity of some points in the speckle pattern \cite{PRTG}. For a QR-PUF, the challenges and the outcomes are quantum states \cite{BS}.
In both cases, however, challenges, outcomes and responses are stored in the CRT as digital binary strings, and the responses are used as authentication keys.

There are two different layers involved in this protocol: a physical one, where the actual $\text{(QR-)}$ PUF acts as a physical evolution from input systems to output systems, and a mathematical one, where a binary challenge string (which should represent the information on how to implement the input system) is mapped into an outcome string which is post-processed into a response string.

To deal with the two different layers, we denote as \emph{challenges} (\emph{outcomes}, \emph{responses}) the strings in the mathematical layer, and as \emph{challenge states} \footnote{This term clearly comes from quantum physics, where it is used to describe a vector in a Hilbert space. We will use the term \emph{classical state} in this article meaning a classical physical quantity, either scalar or vectorial.} (\emph{outcome states}, \emph{response states}) the implementations in the physical layer. 

\begin{figure*}[htb]
\centering
\includegraphics[width=1.0\textwidth]{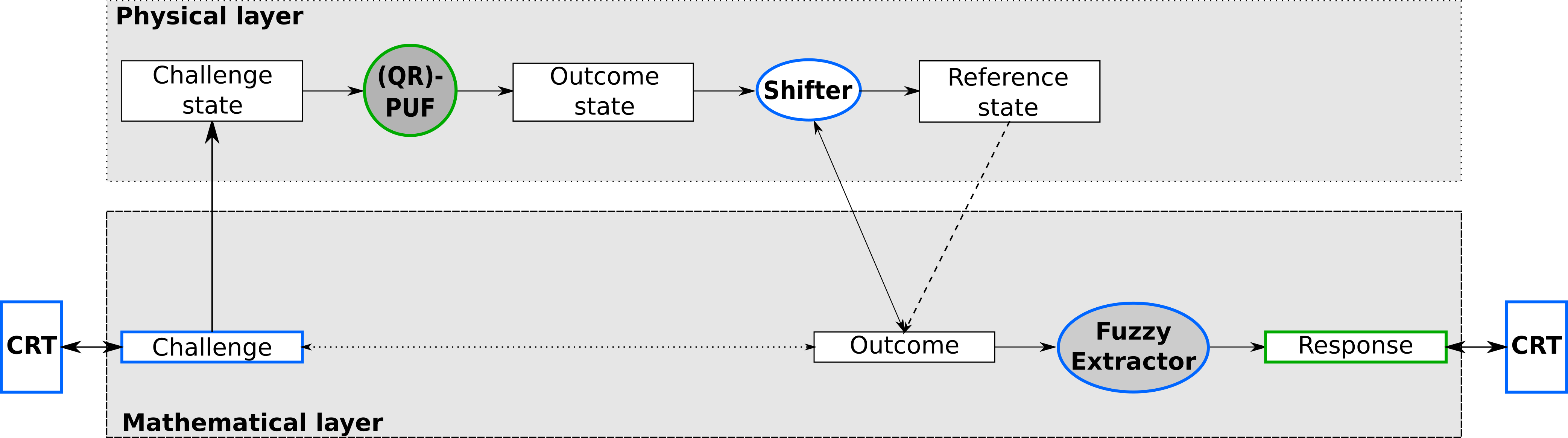}
\caption{A scheme of the two layers, the mathematical one (where the cryptographic protocol takes place) and the physical one (where the $\text{(QR-)}$ PUF acts). In the physical layer a challenge state is prepared according to the information of the challenge (mathematical layer) and then the $\text{(QR-)}$ PUF transforms it into an outcome state. The state-dependent \textit{shifter} (see Secs. \ref{subsec:cenr} and \ref{subsec:qenr}) maps the outcome state to a \textit{reference state}. The outcome in the mathematical layer contains the information about the implementation of the shifter and the error in the reference state and is post-processed by the fuzzy extractor to give the response. Challenges and responses are stored into (enrollment stage) or taken from (verification stage) the Challenge-Response Table (CRT). See Secs. \ref{sec:class} and \ref{sec:quant} for a more detailed description.}
\label{fig:sch}
\end{figure*}

This configuration is schematized in fig. \ref{fig:sch}.

\section{Notation}
\label{sec:not}

In the article we will use the following conventions: 
\begin{itemize}
 \item Digital strings, like the challenges and the responses, are denoted by lowercase bold letters, for instance, $\mathbf{x_i}$ and $\mathbf{r_j}$ for the i-th challenge and the j-th response, respectively;
 \item Sets of digital strings are denoted by the calligraphic uppercase letters, e.g. $\mathcal{X}$ and $\mathcal{R}$ for the set of challenges and responses, respectively;
 \item Random variables which take values from given sets are denoted by uppercase italic letters, e.g. $X$ and $R$ for challenges and responses, respectively;
 \item The physical classical states are denoted by the vector symbol (right arrow), for instance, $\vec{x}_i$ and $\vec{r}_j$ for the i-th challenge state and the j-th response state, respectively;
 \item The physical quantum states are denoted by the usual ket notation, for instance, $\Ket{x_i}$ and $\Ket{r_j}$ for the i-th challenge state and the j-th response state, respectively;
 \item Maps are denoted by uppercase letters with a circumflex accent, e.g. $\hat{P}$ or $\hat{\Pi}$. In particular, the Latin letters are used for maps between strings and the Greek ones for maps between states.
\end{itemize}

\section{Classical PUF}
\label{sec:class}

The realization of a challenge state may involve several different steps, each of them with different experimental complexity.

Each step involves devices with a limited, even though possibly large, number of different configurations and such configurations can be used to parametrize the experimental system, resulting in our ability to formalize the challenges through discrete variables.
A challenge is therefore defined as the binary string $\bf x_i$ of length $n$ representing the configuration which realizes a given challenge state $\vec{x}_i$.
\subsection{Enrollment}
\label{subsec:cenr}

At the start of the enrollment stage, the PUF Certifier selects $N\leq 2^n$ different challenges ${\bf x_i}\in \mathcal{X} \subseteq \{0,1\}^n$, where $\mathcal{X}\subseteq\{0,1\}^n$ is the set of all chosen challenges and $|\mathcal{X}|=N$. In fact, if a challenge consists of $n$ bits, the total possible number of challenges is $2^n$. However, in practice, certain challenges could represent states which are impossible or hard to implement or they do not lead to a set of distinguishable responses. 

For security purposes, the set of challenges $\mathcal{X}$ has to be \textit{uniform}, i.e. $\hat{S}(X)=|\mathcal{X}|$, where $X$ is the random variable defined on the set $\mathcal{X}$ and $\hat{S}(X)$ is the Shannon entropy of $X$. An adversary should not be able to characterize the set of challenges by studying some of them. The Certifier is free to discard some challenges from $\mathcal{X}$ if he finds correlations in them. This affects the number $N$ of challenges and has to be quantified for given experimental implementations.

Each $\bf x_i\in\mathcal{X}$ represents a challenge state $\vec{x}_i$ which can be experimentally realized and sent to the PUF, that acts as a deterministic function $\hat{\Pi}$. Due to its complex structure, any attempt to give a full description of it should be unfeasible, even for the Certifier itself.
For a given challenge state $\vec{x}_i$, $\hat{\Pi}(\vec{x}_i)= \vec{y}_i$, where $\vec{y}_i$ is denoted as \emph{outcome state}.

The Certifier needs to map the outcome state into an outcome string, taking into account both the distribution of the outcome states and any error which may have occurred due to noise or wrong implementation of the experimental system. To do this, we introduce the notion of a \emph{shifter}.

For each outcome state $\vec{y}_i$, let $\hat{\Omega}_i$ be a state-dependent operation, which maps $\vec{y}_i$ into a \textit{reference state}, denoted by $\vec{0}$, equal for all outcome states. For $N$ outcome states $\vec{y}_i$, we obtain a set of $N$ shifters $\hat{\Omega}_i$.
The importance of using the shifters will be more clear when we discuss QR-PUFs. The shifters simplify the error verification process, as each expected outcome is identical. 

Some devices ascribable to shifters have been used in some PUF implementations: consider, for instance, the optical PUF \cite{PRTG}, where a laser beam (challenge state) is transformed into a complex speckle pattern (outcome state). In this scenario, it has been proposed \cite{GHMSP} to use spatial light modulators to transform every speckle pattern into a plane wave, which then is focused into a single point (the reference state). Only if the pattern is the expected one this happens, otherwise, the outcome state is mapped into another speckle pattern.
Shifters can be designed also for other PUFs, depending on which physical quantities are implied in the outcome states. If the outcome state is already a binary value (like in the \emph{SRAM PUF} \cite{GKST}) the reference state can be the bit $0$ and the shifters can be realized by a gate implementing either the identity or a bit flip operation, depending on the expected outcome state. Whenever an outcome is determined by the frequency of a signal (like in a \emph{ring oscillator PUF} \cite{BNCF}), the shifters can be passband filters.

The Certifier can implement the corresponding shifter for every outcome state, since he can characterize $\hat{\Pi}(\vec{x}_i)$, possibly repeating the PUF evaluation for the same challenge state $\vec{x}_i$, to find a $\hat{\Omega}_i$ such that $\hat{\Omega}_i\big(\hat{\Pi}(\vec{x}_i)\big)=\vec{0}$. 

We define $\vec{o}_i:=\hat{\Omega}_i\big(\hat{\Pi}(\vec{x}_i)\big)$. While in the enrollment stage, or in a noiseless verification stage, $\vec{o}_i=\vec{0}$ by definition, in reality $\vec{o}_i$ will contain errors. 
This error is mapped into the Hamming weight, i.e. the number of bits that are different from $0$, of a classical string $\mathbf{o_i}$, i.e. $\mathbf{o_i}=\mathbf{0}_{l_o}=00\dots0$ if and only if $\vec{o}_i=\vec{0}$. The string has a length $ l_o$, dependent on the experimental implementation of the shifter. In the aforementioned example of an optical PUF, the plane wave is focused onto an analyzer plane with a pinhole. If $\vec{o}_i=\vec{0}$ the light passes through this pinhole, and a detector will click. Therefore the intensity of the light on the analyzer plane outside the pinhole can be used to find $\mathbf{o_i}$, and the resolution of the analyzer plane determines the length $l_o$.

The shifters convey information about the distribution of the outcome states (as they are designed on them) and therefore indirectly about the PUF.
We can represent this information in terms of binary strings in the mathematical layer, just as we did for challenge states. The shifters are implemented by an experimental device (or a collection of them) with a limited number of configurations, each one of them implementing a different $\hat{\Omega}_i$. 
Parametrizing such configurations, we map each shifter $\hat{\Omega}_i$ in a string ${\bf w_i}\in\mathcal{W}\subseteq\{0,1\}^{l_w}$. This string is exact, because it represents only the correct implementation of the shifter, without taking into account any noise.
The length $l_w$ depends on the entropy of the shifters and, consequently, on the outcome states (for some implementations, methods to analyze such an entropy have been derived \cite{TSSAO, RSGD}). The entropy of $\mathcal{W}$ has to be studied also to verify the presence of non-uniformity, i.e. correlations between different outcomes or between challenges and corresponding outcomes. This entropy affects the \emph{unclonability} of the PUF (see Sec. \ref{sec:prop}). 

The two strings ${\bf o_i}$ and ${\bf w_i}$ convey two different aspects of the outcome state. In fact, ${\bf o_i}$ gives information about the error only, without distinguishing different outcomes. Instead, ${\bf w_i}$ gives information about the distribution of the outcome states, but not about errors (even a single bit flip of ${\bf w_i}$ changes it into ${\bf w_{j\neq i}}$). 

We combine ${\bf o_i}$ and ${\bf w_i}$ by defining as \emph{outcome} a string $\mathbf{y_i}$ of length $l= l_w+l_o$, such that
\begin{equation}
\mathbf{y_i}=\bf w_i\,\|\,o_i\,,
\end{equation}
where $\|$ is the concatenation of strings.

\begin{figure}[htb]
\centering
\includegraphics[width=0.4\textwidth]{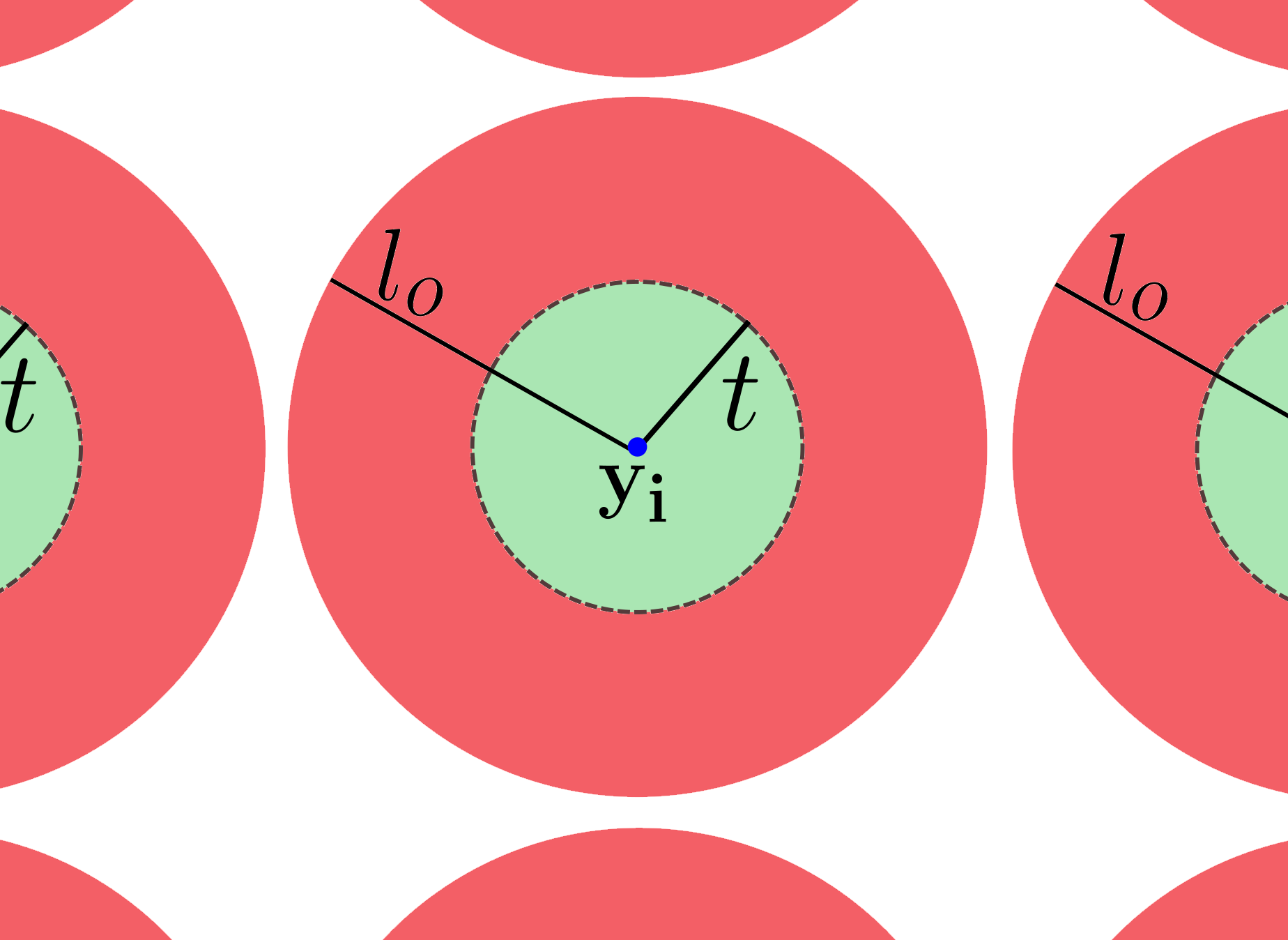}
\caption{Graphic representation of the set $\mathcal{Y}$, according to Eq. \eqref{setY}. The centers of the circles represent the noiseless outcomes ${\bf y_i}={\bf w_i}\|{\bf 0}_{l_o}$ for different ${\bf w_i}\in\mathcal{W}$, while every point in the corresponding outer circles, of radius $l_o$, represents a noisy version of them. Between different outcomes, including the noisy versions, there is no overlap, because $\bf w_i\neq w_j$ for $i\neq j$. A fuzzy extractor can correct $t<l_o$ bit errors, i.e. the outcomes inside the inner circles.}
\label{fig:setY}
\end{figure}

We designate $\mathcal{Y}\subseteq\{0,1\}^l$ as the set of all outcomes, including all possible noisy versions. Explicitly, 
\begin{equation}
\label{setY}
 \mathcal{Y}=\left\{\mathbf{y_i}=\mathbf{w_i}\,\|\,\mathbf{o_i},\, \mathbf{w_i}\in\mathcal{W},\, \mathbf{o_i}\in\{0,1\}^{l_o} \right\}\, ,
\end{equation}
and $|\mathcal{Y}|=2^{l_o}\,N$ (see fig. \ref{fig:setY} for a graphic representation of the set $\mathcal{Y}$).
Moreover we define a function $\hat{P}:\mathcal{X}\rightarrow\mathcal{Y}$, associating each challenge with the corresponding outcome, i.e. $\hat{P}(\bf x_i)=y_i$.

The outcome string, being noisy and not uniformly distributed, cannot be used directly as a response. The most common way to post-process it is through a \emph{fuzzy extractor} \cite{DORS}, which is a combined error correction and privacy amplification scheme:
\begin{definition}
Let $\{0,1\}^\star$ be the \textit{star closure} of $\{0,1\}$, i.e. the set of strings of arbitrary length:
\begin{equation}
\{0,1\}^\star=\bigcup_{i \ge 0 }\{0,1\}^i\, ,
\end{equation}
where $\{0,1\}^0=\emptyset$ is the empty set.
Let $\hat{H}({\bf y_i,y'_i})$ be the \textit{Hamming distance} between $\bf y_i$ and $\bf y'_i$, i.e. the Hamming weight of $\bf y_i+y'_i$ and $ s:= -\log \left(\max_k p_k\right)$ be the \emph{min-entropy} of a probability distribution $p=\Set{p_k}$. Furthermore, given two probability distributions $p_A$, $p_B$, associated to discrete random variables $A,B$ with the same domain $\mathcal{C}$, let $\hat{D}_S(p_A, p_B)$ be the \emph{statistical distance} between $p_A$ and $p_B$,  i.e.
\begin{equation}
\label{statdist}
\hat{D}_S(p_A,p_B):=\frac{1}{2}\,\sum_{c\in\mathcal{C}}\,\left|Pr(A=c)-Pr(B=c)\right|\, .                                                                                                                                                                                                                                                                                                                                              \end{equation}

A $(\mathcal{Y},s,m,t,\epsilon)$-\emph{fuzzy extractor} is a pair of random functions,
the \emph{generation function} $\hat{G}$ and the \emph{reproduction function} $\hat{R}$, with the following properties:
\begin{itemize}
\item $\hat{G}:\mathcal{Y}\rightarrow\{0,1\}^m\times\{0,1\}^\star$ on input $\bf y_i\in\mathcal{Y}$ outputs an extracted string ${\bf r_i}\in\mathcal{R}\subseteq\{0,1\}^m$ and a \emph{helper data} ${\bf h_i}\in\mathcal{H}\subseteq\{0,1\}^\star$. While $\bf r_i$ has to be kept secret, $\bf h_i$ can be made public (it can even be physically attached to the PUF);
\item $\hat{R}:\mathcal{Y}\times\mathcal{H}\rightarrow\{0,1\}^m$ takes an element $\bf y'_i\in\mathcal{Y}$
and a helper string $\bf h_i\in\mathcal{H}$ as inputs.
The \emph{correctness property} of a fuzzy extractor guarantees that if $\hat{H}({\bf y_i,y'_i})\leq t$ and $({\bf r_i,h_i})=\hat{G}(\bf y_i)$, then $\hat{R}(\bf y_i')=r_i$;
\item The \emph{security property} guarantees that for any probability distribution on $\mathcal{Y}$ of min-entropy $s$,
the string $\bf r_i$ is nearly uniform even for those who observe $\bf h_i$: i.e. if $({\bf r_i,h_i})=\hat{G}\bf(y_i)$, then 
\begin{equation}
\hat{D}_S(p_{RH}, p_{UH})\leq\epsilon\, ,
\end{equation}
where $p_{RH}$ ($p_{U\! H}$) is a joint probability distribution for $\bf r_i\in\mathcal{R}$ (for a uniformly distributed variable on $m$-bit binary strings) and $\bf h_i\in\mathcal{H}$.
\end{itemize}
\end{definition}

The generation function of a fuzzy extractor is used, in the enrollment stage, to transform the outcome $\bf y_j$ into the uniformly distributed $\bf r_i$, that is the final \emph{response}. We will see later that, in the verification stage, the reproduction function is used on a noisy version of the outcome to generate the same response.

\begin{figure}[htb]
\centering
\includegraphics[width=0.4\textwidth]{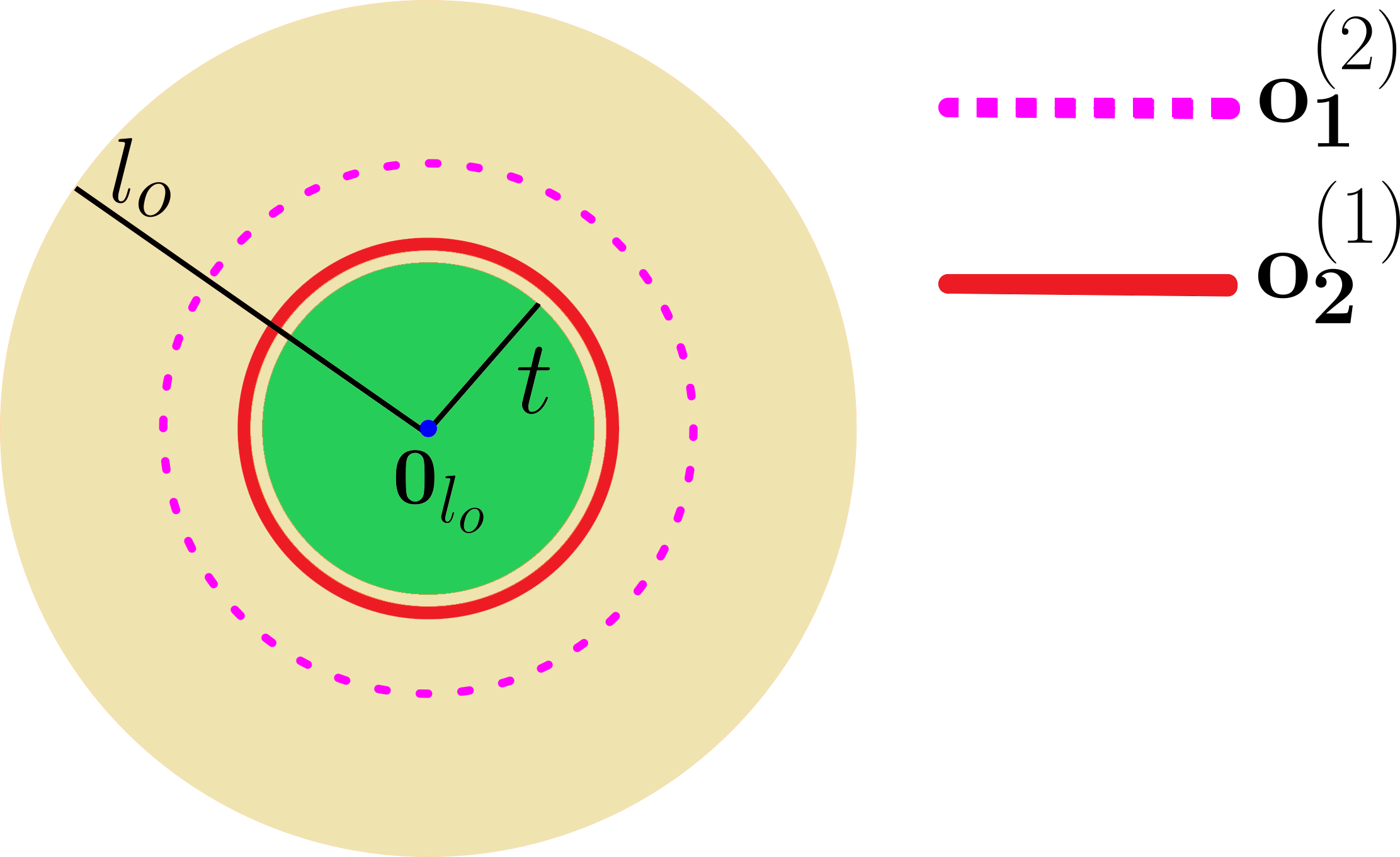}
\caption{Graphic representation of the choice of $t$ for $N=2$ challenge-response pairs. The circle represents both $\bf o_1$ and $\bf o_2$, indipendently from $\bf w_1$ and $\bf w_2$. The center of the circle represent the noiseless cases ${\bf o_1}={\bf o_2}={\bf 0}_{l_o}$ and all the noisy cases lie in a circle of radius $l_o$. The errors ${\bf o}_{\bf 1}^{(2)}$ and ${\bf o}_{\bf 2}^{(1)}$ define two rings and $t$ is chosen smaller than the radius of the smaller one (in our case ${\bf o}_{\bf 2}^{(1)}$).}
\label{fig:overlap}
\end{figure}

The Certifier selects a fuzzy extractor by knowing $\mathcal{Y}$ and its min-entropy $s$, and choosing $t$ such that the fuzzy extractor uniquely maps a given outcome into a response, without collisions: due to noise or an erroneous experimental setup, a challenge state $\vec{x}_i$ can be implemented as a state which is closer to $\vec{x}_j$, for $i\neq j$. 
The error ${\bf o}_{\bf i}^{(j)}$ associated to $\hat{\Omega}_i\big(\hat{\Pi}(\vec{x}_j)\big)$ for $i\neq j$, must be uncorrectable: the Certifier has to choose a maximum allowed error $t<l_o$ smaller than the minimum Hamming weight of ${\bf o}_{\bf i}^{(j)}$, over all $i\neq j$ (see Fig. \ref{fig:overlap}). 

There is a trade-off between $t$ and the entropy of the shifters: a high entropy, associated to a longer length $l_w$ of $\bf w_i$, is equivalent to similar states with a small error in case of a wrong implementation, and $t$ has to be chosen low.
The Certifier may decide to delete challenge-response pairs from the Challenge-Response Table, in order to choose a higher $t$ and increase the resistance of the PUF against the noise. 

For practical purposes we define two functions $\hat{G}_R$ and $\hat{G}_H$ such that
\begin{equation}
\hat{G}(\cdot)=(\hat{G}_R(\cdot),\hat{G}_H(\cdot))\, ,
\end{equation}
and therefore ${\bf r_i}=\hat{G}_R(\bf y_i)$ and ${\bf h_i}=\hat{G}_H\bf(y_i)$ for $\bf y_i\in\mathcal{Y}$.
Moreover, we define the function $\hat{F}_E$ to be the function mapping each challenge to the respective response in the enrollment stage, i.e.
\begin{equation}
\label{eq:fe}
 \hat{F}_E(\cdot):=\hat{G}_R(\hat{P}(\cdot))\, ,
\end{equation}
for $\bf x_i\in\mathcal{X}$ and therefore ${\bf r_i}=\hat{F}_E ({\bf x_i})$.

Summarising, during the enrollment stage the Certifier creates a set of $N$ challenges $\mathcal{X}\in\{0,1\}^n$ and a set of $N$ responses $\mathcal{R}\subseteq\{0,1\}^m$
\begin{equation}
\mathcal{R}=
\Set{{\bf r_i}\in\{0,1\}^m\,|\,{\bf r_i}=\hat{F}_E({\bf x_i});\quad {\bf x_i}\in\mathcal{X}}\, .
\end{equation}

They are stored into the Challenge-Response Table (CRT) together with
\begin{itemize}
 \item the set of $N$ strings $\bf w_i$ representing how to set the shifter operator to get the correct outcome;
 \item the parameters of the fuzzy extractor;
 \item the (possibly public) set of helper data $\mathcal{H}\subseteq\{0,1\}^\star$, i.e.
\begin{equation}
\mathcal{H}=
\Set{{\bf h_i}\in\{0,1\}^\star\,|\,{\bf h_i}=\hat{G}_H(\hat{P}({\bf x_i}));\quad {\bf x_i}\in\mathcal{X}} \, .
\end{equation}
\end{itemize}

The Challenge-Response Table is given to Alice and the PUF to Bob, concluding the enrollment stage.

\subsection{Verification}
\label{subsec:cver}

In the verification stage, Bob declares his identity and allows Alice to (remotely) interact with his PUF. Alice, equipped with the CRT, retraces the steps made by the Certifier in the enrollment stage.

She picks up a randomly selected challenge ${\bf x_j}\in\mathcal{X}$ (for which she knows the response ${\bf r_j}=\hat{F}_E({\bf x_j})$) and prepares the challenge state $\vec{x}_j$. The PUF transforms $\vec{x}_j$ into the outcome state $\hat{\Pi}(\vec{x}_j)$. At this point, Alice tunes the shifter $\hat{\Omega}_j$, according to the CRT and evaluates $\hat{\Omega}_j\big(\hat{\Pi}(\vec{x}_j)\big)$.

After the use of the PUF and the shifter, she may obtain a noisy version of $\vec{y}_j$, because of noise or a wrong preparation of the challenge state. Moreover, the noise could come from the PUF not being the original one, if an adversary Eve is impersonating Bob. 

We call this noisy version $\vec{y'}_j= \hat{\Pi}^{(e)}(\vec{x}_j)$. In that case $\hat{\Omega}_j(\vec{y'}_j)\neq\vec{0}$, which leads to $\mathbf{o'_j}\neq \mathbf{0}_{l_o}$ such that ${\bf y'_j}={\bf w_j\,\|\,o'_j}=\hat{P}^{(e)}({\bf x_j})$ is different from the ${\bf y_j}$ obtained by the Certifier in the enrollment stage.

The outcome is then post-processed by the reproduction function of the fuzzy extractor that was used in the enrollment stage, so Alice collects $\mathbf{z_j}:= \hat{F}_V(\bf x_j)$, where the function $\hat{F}_V$ represents the map between the challenges and the corresponding responses in the verification stage, i.e.
\begin{equation}
\label{eq:fv}
\hat{F}_V:=\hat{R}\big(\hat{P}^{(e)}(\mathbf{\cdot}), \hat{G}_H(\hat{P}(\mathbf{\cdot}))\big) \, ,
\end{equation}
for $\bf x_j\in\mathcal{X}$.

The claimed response $\bf z_j$ is compared with the one in the CRT: if $\bf z_j=r_j$, Bob is authenticated, otherwise the protocol fails.

\section{QR-PUF}
\label{sec:quant}
The authentication scheme for Quantum Readout PUFs follows the structure of the classical scheme (see Sect. \ref{sec:class}) and still uses classical challenges, responses and fuzzy extractors in the mathematical layer. However, the implementation of the challenge states and outcome states in the physical layer is done via quantum states.
At the moment, the only classical PUF which was extended to a QR-PUF is an optical PUF \cite{BS, GHMSP}, for which there are some studies on side-channel attacks \cite{SMP, BS13, YGLZ}.

In this work, we study discrete qubit states, but our approach could also be generalized to continuous-variable $\text{(QR-)}$ PUFs \cite{ND, GN}.

Let us assume to work with $\lambda$ qubits, so challenge states are elements of the Hilbert space $\mathbb{C}^{2^\lambda}$. We also assume that each qubit can be in a finite number of states. Like in the classical case, we can parametrize the configurations of the experimental system that implements the challenge states, to obtain a set $\mathcal{X}$ of classical challenges. Let us denote the length of such strings by $n$, to match the case of classical PUFs. 

Here the challenge states are quantum, therefore challenge states will be represented by $\ket{x_i}$.
Our QR-PUF will be described in an idealized way, as unitary operation acting on a pure state to produce another pure state. In reality, this process will introduce noise: in our framework, this will be taken into account in the transition from the outcome state to the outcome string.

\subsection{Enrollment}
\label{subsec:qenr}

Since not all states are implementable, or they do not lead to distinguishable responses, the Certifier selects $N\leq 2^{n}$ challenges ${\bf x_i}\in\mathcal{X}\subseteq\{0,1\}^n$, 
where $\mathcal{X}$ is implemented by a set of nonorthogonal states $\Set{\Ket{x_1},\dots,\Ket{x_N}}\in\mathbb{C}^{2^\lambda}$.

The nonorthogonality is expected to be a crucial condition, since, as a consequence of the no-cloning theorem \cite{WZ}, there does not exist a measurement which perfectly distinguishes nonorthogonal states. We expect that this enhances the security of QR-PUFs compared to classical PUFs since an adversary could gain only a limited amount of information about the challenge and the outcome states. 

In this work we consider separable challenge states $\Ket{x_i}$, so $\ket{x_i}=\bigotimes_{k=1}^\lambda\ket{x_{ik}}$ and we can deal with single qubit states $\ket{x_{ik}}$. The procedure can be generalized to other challenge states. The qubit states can be written in terms of some complete orthonormal basis, which we denote as $\Set{\Ket{0},\,\Ket{1}}$:
\begin{equation}
\label{chalstat}
\ket{x_{ik}}= \cos \theta_{ik} \, \ket{0} + e^{i \varphi_{ik}} \sin \theta_{ik} \,\ket{1}\, ,
\end{equation}
where $\theta_{ik}\in[0,\pi]$ and $\varphi_{ik}\in[0,2\pi]$.

The Certifier sends all states to the QR-PUF, collecting the outcome states. The QR-PUF is formalized as a $\lambda$-fold tensor product of single-qubit unitary gates $\hat{\Phi}=\bigotimes_{k=1}^\lambda \hat{\Phi}_k$. Despite its form being unknown, it can be parametrized by \cite{ZK}:
\begin{equation}
\label{unitmat}
\hat{\Phi}_k(\omega_k,\psi_k,\chi_k)=
\begin{pmatrix}
  e^{i \psi_k}\cos \omega_k & e^{i \chi_k}\sin \omega_k \\
  -e^{-i \chi_k}\sin \omega_k & e^{-i \psi_k}\cos \omega_k
\end{pmatrix}\, ,
\end{equation}
with random parameters $\psi_k, \chi_k \in [0, 2\pi]$ and $ \omega_k \in \left[0, \frac{\pi}{2}\right]$.
The outcome state is then $\ket{y_i}=\bigotimes_{k=1}^\lambda\ket{y_{ik}}$, where
\begin{equation}
\label{outstat}
\begin{split}
&\ket{y_{ik}}=\,\hat{\Phi}_k \ket{x_{ik}}\\
&=\begin{pmatrix}
  e^{i \psi_k}\cos \omega_k \cos \theta_{ik} + e^{i (\chi_k+\varphi_{ik}) }\sin \omega_k \sin \theta_{ik} \\
  -e^{-i \chi_k}\sin \omega_k \cos \theta_{ik} + e^{i(\varphi_{ik}- \psi_k)}\cos \omega_k \sin \theta_{ik}
\end{pmatrix}\, .
\end{split}
\end{equation} 

\begin{figure*}[htb]
\centering
\includegraphics[width=1\textwidth]{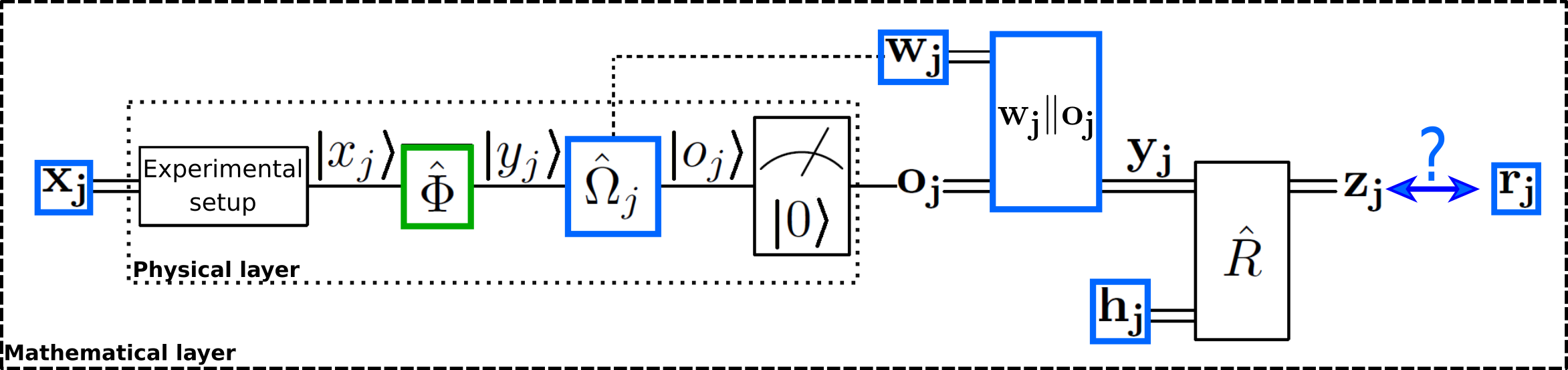}
\caption{A scheme for the verification stage for QR-PUFs, as described in Sec. \ref{subsec:qver}. Bob provides the QR-PUF ($\hat{\Phi}$, here enclosed in a green box) and Alice uses quantities stored in the Challenge-Response Table (here enclosed in blue boxes) to evaluate a response ${\bf z_j}$ for a challenge ${\bf x_j}$. Authentication succeeds if ${\bf z_j}={\bf r_j}$, where ${\bf r_j}$ is the response stored in the CRT. The verification stage for classical PUFs (as described in Sec. \ref{subsec:cver}) can be obtained by substituting in the physical layer (the inner box) quantum states and operators with classical states and operators, and by leaving the mathematical layer (outer box) unchanged.}
\label{fig:qPUF}
\end{figure*}
Like in the classical case, the Certifier can design a state-dependent shifter, that performs a tensor product of unitary transformations, $\hat{\Omega}_{i}=\bigotimes_{k=1}^\lambda\hat{\Omega}_{ik}$, each one of them mapping a specific qubit state to the reference state $\ket{0}=(1,0)^T$. This operation is indeed unitary, because for $\ket{y_{ik}}= \cos\alpha_{ik}\Ket{0}+e^{i\beta_{ik}}\sin\alpha_{ik}\ket{1}$, it holds that $\hat{\Omega}_{ik} \ket{y_{ik}} =\Ket{0}$ for
\begin{equation}
\label{shifdef}
\hat{\Omega}_{ik} =
\begin{pmatrix} 
 \cos\alpha_{ik} & e^{-i\beta_{ik}}\sin\alpha_{ik}\\
 e^{i\beta_{ik}}\sin\alpha_{ik} & -\cos\alpha_{ik}
\end{pmatrix}\, ,
\end{equation}
which verifies $\hat{\Omega}_{ik}\,\hat{\Omega}^\dagger_{ik}=\hat{\Omega}_{ik}^\dagger\,\hat{\Omega}_{ik}=\mathbb{I}$, where $\mathbb{I}$ is the identity operator.
The Certifier can implement $\hat{\Omega}_{i}$ for each $\hat{\Phi}\Ket{x_i}$ because he can repeat the experiment and characterize each outcome state by performing quantum state tomography or, as we work with pure states, compressed sensing \cite{GLFBE}.

Instead of having to change the single-qubit measurement basis for each qubit and each challenge, by applying the suitable shifter it is now possible to use the basis $\set{\Ket{0}, \Ket{1}}$ for all qubits of all challenges.

By definition of $\hat{\Omega}_{ik}$, if there is no error, we will measure for every qubit the state $\ket{0}$, and the results of the measurement form a string of length $\lambda$ made by all zeros, ${\bf o_i=0}=00\dots0$. If there is some error, which in the quantum case is introduced by either the environment or an adversary, the Hamming weight of $\bf o_i$ will give us an estimate of it.  

Like in the classical case, we can parametrize the experimental system that implements the shifters in terms of the (discrete) configuration it must assume to implement a specific $\hat{\Omega}_i$. Therefore, a given $\hat{\Omega}_i$ is represented by a classical string $\bf w_i\in\mathcal{W}$ of length $ l_w$.

We again define as \emph{outcome} a classical string $\mathbf{y_i}$ of length $l=l_w+\lambda$, given by:
\begin{equation}
\mathbf{y_i}=\bf w_i\,\|\,o_i\, ,
\end{equation}
where $\|$ is the concatenation of strings.

We also define a set $\mathcal{Y}$ like in Eq. \eqref{setY} and a function $\hat{P}:\mathcal{X}\rightarrow\mathcal{Y}$ mapping every challenge to the corresponding outcome.

At this point, like for classical PUFs, the Certifier fixes the correctable amount of noise $t<l_o$ and selects a fuzzy extractor $(\hat{G}, \hat{R})$, able to correct $t$ errors and to generate a uniformly distributed response, according to the distribution of the outcome states and the entropy of the set of outcomes. The non-orthogonality of the challenge states affects $t$: when a wrong challenge state is implemented, its \textit{fidelity} with the correct one is preserved by the QR-PUF and the shifter, since they are unitary maps, and influences the results of the measurement. The maximum correctable error $t$ has to be chosen lower than the error produced by wrong implementations, which becomes small for highly non-orthogonal challenges. The Certifier may decide to delete challenge-response pairs from the Challenge-Response Table, in order to choose a higher $t$ and increase the resistance of the QR-PUF against the noise. However, this reduces the overall non-orthogonality of the quantum states, thus improving Eve's ability to distinguish them. Such a trade-off will be discussed again in the following sections.

The generation function of a fuzzy extractor generates a uniformly distributed response $\bf r_i \in\mathcal{R}$, together with a public helper data $\bf h_i\in\mathcal{H}$.
Again we have:
\begin{equation}
\hat{G}(\cdot)=(\hat{G}_R(\cdot),\hat{G}_H(\cdot))\, ,
\end{equation}
and
\begin{equation}
{\bf r_i}=\hat{G}_R({\bf y_i}), \quad\forall \,{\bf y_i}\in\mathcal{Y}\, .
\end{equation}

We define a function $\hat{F}_E(\cdot):=\hat{G}_R(\hat{P}(\cdot)):\mathcal{X}\rightarrow\mathcal{R}$ mapping each challenge to the corresponding response, representing the action of the QR-PUF in the enrollment stage.
Like for classical PUFs, challenges, responses and other information are stored in the Challenge-Response Table, which is given to Alice, while the QR-PUF is given to Bob.
\subsection{Verification}
\label{subsec:qver}
In the verification stage Bob allows Alice to (remotely) interact with his $\text{QR-}$PUF. She selects randomly a challenge $\bf x_j\in\mathcal{X}$ and prepares $\Ket{x_j}$.

Using the QR-PUF with the challenge state $\Ket{x_j}$, Alice may obtain $\ket{y'_j}$, different from the expected $\ket{y_j}$, because of noise or an erroneous implementation of the system or the action of a malicious intruder. Then Alice applies $\hat{\Omega}_{j}$ and measures each qubit state in the basis $\Set{\Ket{0},\Ket{1}}$, obtaining $\bf o'_j$ and hence the outcome $\mathbf{y'_j}=\bf w_j\,\|\,o'_j$.
While in the ideal noiseless case $\mathbf{o'_j}={\bf 0}_{l_o}$, here we may measure some state $\ket{1}$ for some qubits, therefore $\mathbf{y'_j}$ could be different from the $\bf y_j$ obtained by the Certifier in the enrollment stage.

The outcome is then post-processed by the reproduction function of the fuzzy extractor that was used in the enrollment stage, so Alice collects $\mathbf{z_j}:= \hat{F}_V(\bf x_j)$, where the function $\hat{F}_V$ is defined like in the classical case, $\hat{F}_V(\mathbf{\cdot}):=\hat{R}\big(\hat{P}^{(e)}(\mathbf{\cdot}), \hat{G}_H(\hat{P}(\mathbf{\cdot}))\big)$.

Authentication succeeds if $F_E({\bf x_j})=F_V({\bf x_j})$.
The verification stage is schematized in fig. \ref{fig:qPUF}.

\section{Properties and formalization}
\label{sec:prop}
In this section, we will analyze the properties of $\text{(QR-)}$ PUFs. As we have seen, both PUFs and QR-PUFs can be represented by a classical pair of functions $\hat{F}=(\hat{F}_E,\hat{F}_V)$ that describe the map between challenges and responses in the enrollment ($\hat{F}_E$, see Eq. \eqref{eq:fe}) or verification ($\hat{F}_V$, see Eq. \eqref{eq:fv}) stage. We will keep the same formalism for both PUFs and QR-PUFs, to allow our framework to compare them, but we will also specify the practical differences.

We have seen that the noise can be a problem which can lead to false rejection in the protocols.
Therefore it is important to characterize and quantify the amount of noise of a $\text{(QR-)}$ PUF, which is connected to the \emph{robustness} of a $\text{(QR-)}$ PUF.
We take the definition of this concept from \cite{AMSSW}, adapting it to our framework and our formalism.
\begin{definition}
Let us consider a $\text{(QR-)}$ PUF $\hat{F}$ with a set of challenges $\mathcal{X}$, where $|\mathcal{X}|=N$.

$\hat{F}$ is $\rho$-\emph{robust} with respect to $\mathcal{X}$ if $\rho\in[0,1]$ is the greatest number for which:
\begin{equation}
\frac{1}{N}\sum_{i=1}^{N}\,
Pr\{\hat{F}_V({\bf x_i})=\hat{F}_E({\bf x_i})\} \geq \rho\, .
\end{equation}

$\rho$ is called the \emph{robustness} of the $\text{(QR-)}$ PUF with respect to $\mathcal{X}$.
\end{definition}

The robustness represents the average probability that the $\text{(QR-)}$ PUF in the verification stage outputs the correct response, such that the authentication succeeds. So it represents the $\text{(QR-)}$ PUF's ability to avoid false rejections and depends on many factors, e.g. on the average noise of the specific implementation and the parameters of the fuzzy extractor.

Regarding the robustness, we do not expect a significant advantage of QR-PUFs compared to classical PUFs.
Actually, there is the possibility to have a disadvantage, because of the fragility of quantum states and of the necessity of having a low error threshold $t$, as the noise can originate from a possible interaction of an adversary.
Any implementation with QR-PUFs has to pay special care to this issue.

Now we will discuss unclonability, which is the main parameter involved in attacks from an adversary Eve. This concept is also mildly inspired by \cite{AMSSW}, but with marked differences, mainly caused by the need of taking into account QR-PUFs.
In the context of entity authentication with $\text{(QR-)}$ PUF, the purpose of an adversary Eve is to create a clone of a $\text{(QR-)}$ PUF, such that Alice can verify with it a challenge-response pair, falsely authenticating her as Bob.

When we say \emph{clone}, we need to specify if we are talking of a physical or a mathematical one.
A \emph{physical clone} is an experimental reproduction of the $\text{(QR-)}$ PUF. It will have the same physical properties as the original one, even in contexts not involved with the authentication protocol.
The requirement of \emph{physical unclonability} means that a physical clone is technologically or financially unfeasible at the current state of technology.

A mathematical clone, instead, is an object that \emph{simulates} the challenge-response behavior of a $\text{(QR-)}$ PUF. 
In this case, we cannot just state that a mathematical clone is unfeasible, because if there are some correlations between the outcome states, in principle they can be exploited to predict new challenge-response pairs. As mentioned in the introduction, several PUFs have been successfully mathematically cloned. We need to formalize this notion, in order to quantify it for different $\text{(QR-)}$ PUFs.

We assume that Eve cannot directly access the internal structure of the $\text{(QR-)}$ PUF \cite{RSS, RBK}, but only interact with the challenge and the outcome states.
An attack consists of two phases, both carried out during the verification stage of the protocol. We require that the enrollment stage is inaccessible to Eve since this part is performed in the Certifier's lab and it involves the study of the inner structure of the $\text{(QR-)}$ PUF. During the \emph{passive phase}, Eve observes a certain number of successful authentications with the real $\text{(QR-)}$ PUF, collecting as much information as she can. Then, during the \emph{active phase} she designs a clone and gives it to Alice, claiming to be Bob. The attack succeeds if she is authenticated as Bob.

Each interaction affects one challenge-response pair. In this context, there is a crucial difference between PUFs and QR-PUFs. Classical states can be measured without introducing disturbances and can be copied perfectly. Therefore for $q\leq N$ interactions, we can assume that Eve would know exactly $q$ challenge and outcome states, possibly using this information to create a mathematical clone of the PUF.

Instead, a quantum state cannot be copied. Moreover, a quantum measurement cannot perfectly distinguish the states (since they are non-orthogonal) and any measurement can in principle introduce errors, thus potentially making a passive eavesdrop a detectable action. After $q$ interactions, Eve would know less than $q$ challenge and outcome states. This is the main reason for which QR-PUFs have been introduced:  we expect that, concerning unclonability, they can be superior, even far superior, than classical PUFs \footnote{As we mentioned in Sec. \ref{subsec:qenr}, highly non-orthogonal challenge states require a fuzzy extractor with a low correctable error, undermining the robustness of the QR-PUF. Therefore this feature of QR-PUFs must be used carefully, balancing robustness and unclonability.}.

\begin{definition}
Let $\hat F$ be a $\text{(QR-)}$ PUF with a set of challenges $\mathcal{X}$, where $|\mathcal{X}|=N$. Let us suppose that an adversary Eve has $q$ interactions with a $\text{(QR-)}$ PUF in the passive stage of an attack, by observing an authentication protocol between Alice and Bob. With the information she can extract, she prepares a clone $\hat{E}_q$, defined as (see Eq. \eqref{eq:fv} for a comparison) 
\begin{equation}
\label{eq:Eq}
\hat{E}_q(\cdot):=\hat{R}\big(\hat{P}_E (\cdot), \hat{G}_H(\hat{P}(\cdot))\big)\:,
\end{equation}
and gives it to Alice, who selects a challenge $\bf x_i\in\mathcal{X}$ and evaluates $\hat{E}_q({\bf x_i})$.

Then $\hat{E}_q$ is a $ (\gamma,q)$-\emph{(mathematical) clone} of $\hat{F}$ if $\gamma\in[0,1]$ is the greatest number for which
\begin{equation}
\frac{1}{N}\sum_{i=1}^{N} Pr(\hat{E}_q({\bf x_i})=\hat{F}_E({\bf x_i}))\geq \gamma\, .
\end{equation}
\end{definition}

\begin{definition}
A $\text{(QR-)}$ PUF $\hat{F}$ is called $(\gamma,q)$-\emph{(mathematical) clonable} if $\gamma\in[0, 1]$ is the smallest number for which it is not possible to generate a $(\bar{\gamma},q)$ clone of the $\text{(QR-)}$ PUF for any $\bar{\gamma}>\gamma$.

Conversely, a $\text{(QR-)}$ PUF $\hat{F}$ is denoted as $(\delta,q)$-\emph{(mathematical) unclonable} if it is $(1-\delta,q)$-clonable.
\end{definition}
The unclonability of a $\text{(QR-)}$ PUF is therefore related to the average probability of false acceptance.

We could expect to find a relation between the number of interactions $q$ and the unclonability: with a higher knowledge of CRP, it could be expected that Eve will be able to build a more and more sophisticated reproduction of the $\text{(QR-)}$ PUF. Increasing $q$ increases the know-how for making $(1-\delta,q)$-clones with a lower $\delta$. Therefore, fixing the maximum number of uses $q=q^* $ we fix the minimum $\delta=\delta^* $.
So we ensure that for $q<q^* $, the $\text{(QR-)}$ PUF is at least $(\delta^* , q)$-unclonable.

\begin{definition}
A $(\rho,\delta^* ,q^* )$-\emph{secure} $\text{(QR-)}$ PUF $\hat{F}$ is $\rho$-robust, physically unclonable and at least $(\delta^* ,q)$-mathematically unclonable up to $q^* $ uses.
\end{definition}

When manufacturing $\text{(QR-)}$ PUFs several properties, that are typically implementation-dependent, are important \cite{MV}. We believe that the above theoretical definitions of robustness and unclonability are, from a theoretical point of view, the main and most general properties involved in a $\text{(QR-)}$ PUF. They are directly related to the probabilities of false rejection and false acceptance, hence describing the efficiency and the security of the entity authentication protocol. 
They also describe all $\text{(QR-)}$ PUFs independently from their implementation. 

\section{Examples}
\label{sec:ex}
Explicit calculation of the robustness and the unclonability for a particular $\text{(QR-)}$ PUFs strongly depends on its implementation. 
In this section, we illustrate the analysis for simplified examples, starting from idealized, extreme, cases.
\begin{itemize}
 \item Consider a physically unclonable device implementing a true random number generator. An example of that is a QR-PUF based on the shot noise of an integrated circuit. 
This device is 
extremely difficult to simulate (Eve has to try a random guess), but also not robust at all (since it will not generate the same number in the enrollment and the verification). For this device, it holds
\begin{equation}
  \begin{split}
  &\frac{1}{N}\sum_{i=1}^{N}\,Pr\{\hat{F}_V({\bf x_i})=\hat{F}_E({\bf x_i})\}= \frac{1}{N}\, ; \\
  &\frac{1}{N}\sum_{i=1}^{N} Pr(\hat{E}_{q^*}({\bf x_i})=\hat{F}_E({\bf x_i}))= \frac{1}{N}\, .
  \end{split}
\end{equation}
Therefore it is a $(1/N,1-1/N,q^*)$ $\text{(QR-)}$ PUF, for any $q^*$.
 \item Consider a physically unclonable device that outputs a fixed signal ($\vec{0}$ for classical PUFs or $\Ket{0}$ for QR-PUFs) for any input. An example is an optical QR-PUF based on the polarization of light for which a fixed polarizer is used as a shifter: for all outcome states only light waves of a specific polarization would pass though. This device is perfectly robust, but also clonable. It holds
 \begin{equation}
  \begin{split}
  &\frac{1}{N}\sum_{i=1}^{N}\,Pr\{\hat{F}_V({\bf x_i})=\hat{F}_E({\bf x_i})\}= 1\, ; \\
  &\frac{1}{N}\sum_{i=1}^{N} Pr(\hat{E}_{q^*}({\bf x_i})=\hat{F}_E({\bf x_i}))= 1\, .
  \end{split}
 \end{equation}
Therefore the $\text{(QR-)}$ PUF is a $(1,0,q^*)$ $\text{(QR-)}$ PUF, for any $q^*$.
\end{itemize}

These examples are extreme cases, while all $\text{(QR-)}$ PUFs will be somewhere in between. We now focus on an example of QR-PUF, to point out some features of QR-PUFs and some open points.

Let $\hat{F}$ be a QR-PUF implemented by a unitary transformation $\hat{\Phi}$, acting on $\lambda$ qubits, parametrized according to Eq. \eqref{unitmat}, with $\psi_k=\chi_k=0$, i.e.
\begin{equation}
 \hat{\Phi}=\bigotimes_{k=1}^\lambda \hat{\Phi}_k=\bigotimes_{k=1}^\lambda 
 \begin{pmatrix}
 \cos\omega_k & \sin\omega_k \\
 -\sin\omega_k & \cos\omega_k
 \end{pmatrix}\, .
\end{equation}

Consider a scenario in which each challenge state is a separable state of $\lambda$ qubits, $\Ket{x_i}=\bigotimes_{k=1}^\lambda \Ket{x_{ik}}$, and each qubit is in one of four possible states:
\begin{equation}
\label{exqub}
 \ket{x_{ik}}=\ket{x_{ik}^{(\ell)}}:=\cos \left(\frac{\phi^{(\ell)}}{
 2}\right)\Ket{0}+\sin \left(\frac{\phi^{(\ell)}}{2}\right)\Ket{1}\, ,
\end{equation}
where 
\begin{equation}
\label{eq:phi} 
\begin{split}
 &\phi^{(1)}=\phi\, ,\qquad\qquad \phi^{(2)}=-\phi\, ,\\ 
 &\phi^{(3)}=\phi-\pi\, ,\qquad\: \phi^{(4)}=\pi-\phi\, ,
 \end{split}
\end{equation}
for a fixed angle $\phi$. Such challenge states can be parametrized by challenge strings of length $n=2\lambda$: for each qubit, the four possibilities are represented by two bits.

For simplicity of notation, from now on, we drop the indices $i$ and $k$, e.g. we write $\big|x^{(\ell)}\big\rangle:=\big| x_{ik}^{(\ell)}\big\rangle$.
The pairs $\{\Ket{x^{(1)}},\Ket{x^{(3)}}\}$ and $\{\Ket{x^{(2)}},\Ket{x^{(4)}}\}$ are orthogonal, but the overall set is non-orthogonal. 

We assume that the noise can be parametrized as a depolarizing channel, associated to a probability of error $\tilde{p}$ and equal for all qubits. The noisy challenge state reads: 
\begin{equation}
\begin{split}
\tilde{\rho}_x&:=(1-\tilde{p})\Ket{x}\Bra{x}+\tilde{p}\,\frac{\hat{I}}{2} \\ 
&=\left[(1-\tilde{p})\cos^2\left(\frac{\phi^{(\ell)}}{2}\right)+ \frac{\tilde{p}}{2}\right] \Ket{0}\Bra{0}\\
&+\left[ (1-\tilde{p}) \sin\left(\frac{\phi^{(\ell)}}{2}\right) \cos\left(\frac{\phi^{(\ell)}}{2}\right) \right] \left(\Ket{0}\Bra{1}+ \Ket{1}\Bra{0}\right) \\
&+\left[(1-\tilde{p}) \sin^2\left(\frac{\phi^{(\ell)}}{2}\right) +\frac{\tilde{p}}{2}\right]\Ket{1}\Bra{1}\, .
\end{split}
\end{equation}

The shifter needs to map the noiseless outcome state to $\Ket{0}\dots\Ket{0}$. According to Eq.\eqref{shifdef} it can be chosen to be a $\lambda$-fold tensor product of single qubit gates 
\begin{equation}
\begin{split}
\hat{\Omega}= &\cos\left(\frac{\phi^{(\ell)}}{2}-\omega\right)\proj{0}{0}+ \sin\left(\frac{\phi^{(\ell)}}{2}-\omega\right)\proj{0}{1}\\ 
+&\sin\left(\frac{\phi^{(\ell)}}{2}-\omega\right)\proj{1}{0}-\cos\left(\frac{\phi^{(\ell)}}{2}-\omega\right)\proj{1}{1}\, ,
\end{split}
\end{equation}
and it follows:
\begin{equation}
\tilde{\rho}_o:= \hat{\Omega}\,\tilde{\rho}_{y}\,\hat{\Omega}^\dagger= \left(1-\frac{\tilde{p}}{2}\right)\proj{0}{0}+ \left(\frac{\tilde{p}}{2}\right)\proj{1}{1} \, .
\end{equation}

For a single qubit, therefore, the probability of measuring $\Ket{1}$ is $\tilde{p}/2$. 
For a challenge state of $\lambda$ qubits, the average Hamming weight of the string $\bf o_i$ is $\lambda\,\tilde{p}/2$.

Any fuzzy extractor is defined in terms of the maximum number of errors $t$ it can correct. With our error model, we can choose to correct the average error of the system, i.e.  $t=\lceil \lambda\,\tilde{p}/2 \rceil$, where $\lceil \lambda\,\tilde{p}/2 \rceil$ is the least integer greater than or equal to $\lambda\,\tilde{p}/2$. 

However, $t$ and the number $N$ of challenge-response pairs are related since the fuzzy extractor has to uniquely map a given outcome into a unique response, without collisions. 

Consider $\big| x^{(\ell)}\big\rangle$ and $\big| x^{(\ell')}\big\rangle$ ($\ell, \ell'\in \{1,2,3,4\}$ and $\ell\neq \ell'$) and estimate the error if $\big| x^{(\ell)}\big\rangle$ is implemented as the state $\big| x^{(\ell')}\big\rangle$, by evaluating $\hat{\Omega}_\ell \,\hat{\Phi}\big|x^{(\ell')}\big\rangle$.
From
\begin{equation}
\begin{split}
&\Ket{x^{(\ell)}}= \cos \left(\frac{\phi^{(\ell)}}{2}\right)\Ket{0}+\sin \left(\frac{\phi^{(\ell)}}{2}\right)\Ket{1}\, , \\
&\Ket{x^{(\ell')}}= \cos \left(\frac{\phi^{(\ell')}}{2}\right)\Ket{0}+\sin \left(\frac{\phi^{(\ell')}}{2}\right)\Ket{1}\, ,
\end{split}
\end{equation}
it follows
\begin{equation}
\begin{split}
&\hat{\Omega}_\ell\, \hat{\Phi}\Ket{x^{(\ell')}} \\
&=\cos\left(\frac{\phi^{(\ell)}-\phi^{(\ell')}}{2}\right)\Ket{0}+ \sin\left(\frac{\phi^{(\ell)}-\phi^{(\ell')}}{2}\right)\Ket{1}\, .
\end{split}
\end{equation}

Therefore, for this case, the probability of measuring $\ket{1}$ is $\sin^2\big[\big(\phi^{(\ell)}-\phi^{(\ell')}\big)/2\big]$.
 
In table \ref{table1}, the explicit values for all the combinations of the 4 qubit states are listed. In case of wrong implementation, challenges with a large overlap lead to small error weights, while orthogonal challenges lead to big ones. Thus there is a trade-off between the robustness of the QR-PUF and the quantum advantage of using indistinguishable non-orthogonal states. 

\begin{table}[ht]
\centering
\begin{tabular}{ c| c c c c } 
  & $\Ket{x^{(1)}}$ & $\Ket{x^{(2)}}$ & $\Ket{x^{(3)}}$ & $\Ket{x^{(4)}}$ \\ 
\hline $\Ket{x^{(1)}}$ & 0 & $\sin^2\phi$ & 1 & $\cos^2\phi$ \\ 
$\Ket{x^{(2)}}$ & $\sin^2\phi$ & 0 & $\cos^2\phi$ & 1 \\
$\Ket{x^{(3)}}$ & 1 & $\cos^2\phi$ & 0 & $\sin^2\phi$ \\
$\Ket{x^{(4)}}$ & $\cos^2\phi$ & 1 & $\sin^2\phi$ & 0 \\
 \hline
\end{tabular}
\caption{Error induced by implementing the wrong challenge state: the entry in row $\ell$ and column $\ell'$ of the table is the probability of error when applying shifter $\ell$ to state $\ell'$. The parameter $\phi$ is defined in Eq. \eqref{eq:phi}.}
\label{table1}
\end{table}
For any pair of possible challenge states $\Ket{x_i}=\bigotimes_{k=1}^\lambda \ket{x_{ik}}$ and $\Ket{x_j}=\bigotimes_{k=1}^\lambda \ket{x_{jk}}$, the average Hamming weight of the error string $\bf o_i$, obtained by the aforementioned process, is 
\begin{equation}
\begin{split}
\operatorname{err}_{i,j}&:=(n_{12}+n_{34})\sin^2\phi+ (n_{13}+n_{24}) \\
&+(n_{14}+n_{23})\cos^2\phi\, ,
\end{split}
\end{equation} 
where $n_{ab}$ counts how many times $\ket{x_{ik}}=\Ket{x^{(\ell)}}$ when $\ket{x_{jk}}=\Ket{x^{(\ell')}}$ (or viceversa).

If $\operatorname{err}_{i,j}<\lceil \lambda\,\tilde{p}/2 \rceil$, then the Certifier should discard one of the two challenges, either $\bf x_i$ or $\bf x_j$, thus reducing the number $N$ of possible challenge-response pairs. 
After this selection is repeated for all pairs of challenges, the Certifier studies the entropy of the set of shifters, determining the strings $\bf w_i$
and the outcomes $\mathbf{y_i}=\bf w_i\,\|\,o_i$.

The \textit{Canetti's reusable fuzzy extractor} \cite{CFPRS} is able to correct up to $t=(l\ln l/m)$ bits, where $l$ is the length of the outcomes and $m$ the length of the responses. As $l=\lambda+l_w$ is fixed, $m$ has to be adapted to the noise level $\lceil \lambda\,\tilde{p}/2 \rceil$. 
The correctness property of this fuzzy extractor guarantees that an error smaller than $t$ is corrected with probability $1-\tilde{\varrho}$, where 
\begin{equation}
\tilde{\varrho} = \left(1-\left(1-\frac{t}{l}\right)^m\right)^{\xi_1}+\xi_1\xi_2 \,,
\end{equation}
with $\xi_1$ and $\xi_2$ being computational parameters of the fuzzy extrator (in \cite{CFPRS}, to which we refer for a precise explanation, they are called $\ell$ and $\gamma$, respectively).

Then the robustness of this QR-PUF is $1-\tilde{\varrho}$.

Concerning the unclonability, one should relate the amount of information Eve obtains from the (possibly correlated) challenge-response pairs to her ability to create a mathematical clone of the QR-PUF. Unfortunately, there is no general method known to provide this relation. We expect that, for some QR-PUFs, quantum unitary gate discrimination methods \cite{CH} could be used, but this line of research goes beyond the purposes of our work. 
Here, we can show that QR-PUFs prevent Eve to gain too much information about challenges and responses, thus strongly hindering her ability to learn the CRT.

As the optimal global attack on the challenge states is unknown, unless knowing all challenge states, here we consider an attack that acts individually on
qubits. 
In particular, we consider the case for which, on each qubit, Eve can apply a $1\rightarrow 2$ cloning operator, i.e. she can intercept each qubit of a challenge state during an authentication round to produce two (imperfect) copies, one of which is given back to the legitimate parties and the other is kept for herself.

For such a set of states, the optimal cloning tranformation, i.e. the transformation who keeps the highest possible fidelity between the copies and the original states, has been derived \cite{BM01} and for any challenge state $\Ket{x_i}$ and its optimal copy $\rho_i^E$ holds: 
\begin{equation}
\begin{split}
 &F(\Ket{x_{i}}\Bra{x_{i}},\varrho^E_{i} ):=\prod_{k=1}^\lambda\Braket{x_{ik}|\varrho^E_{ik}|x_{ik}}\\
 &=\left(\frac{1}{2}\,\left(1+\sqrt{\sin^4\phi+\cos^4\phi}\right)\right)^\lambda\,.
\end{split}
\end{equation}

For fixed $\lambda$, the minimum value of the fidelity is reached for $\phi=\pi/4$, for which, considering a single qubit, $F=(0.85)$.
Already for $10$ qubits the fidelity drops to $F=(0.20)$, and for $20$ qubits, $F=(0.04)$. 

Thus, Eve is not able to successfully simulate the challenge-response behavior, as she cannot even reconstruct the challenge and outcome states.
Moreover, as the fidelity is preserved by unitary matrices, this result holds also for the expected outcome state $\Ket{y_i}$ and the actual outcome state Alice obtains after challenging the QR-PUF with her (unwittingly altered by the cloning process) challenge state. The noise is too high to be corrected by the fuzzy extractor, thus aborting the authentication protocol and exposing the presence of an intruder.   

For classical PUFs, instead, Eve could perfectly read the challenge and outcome states, without being noticed. This provides an advantage of QR-PUFs compared to classical PUFs in terms of unclonability. However, we also noticed that a high non-orthogonality of the challenges can, in principle, undermine the robustness. The trade-off between the advantages and disadvantages of QR-PUFs (Table \ref{table2}) has to be studied to find secure applications of them.
\pagebreak

\begin{table}[htbp]
\centering
\begin{tabular}{|c|c|}
\hline
\multicolumn{2}{|c|}{QR-PUFs compared to PUFs}      \\ \hline
\textbf{ Advantages} & \textbf{ Disadvantages} \\ \hline
  & \\
\begin{tabular}[c]{@{}c@{}} An adversary cannot \\ copy or distinguish \\ non-orthogonal states. \\ \end{tabular} &
  \begin{tabular}[c]{@{}c@{}}Highly non-orthogonal \\ states reduce the \\ robustness. \\ \end{tabular} \\
  & \\
\begin{tabular}[c]{@{}c@{}}Adversarial measurements\\ on the states introduce \\ detectable disturbances.\end{tabular} &
  \begin{tabular}[c]{@{}c@{}}Quantum states are \\ more fragile than \\ classical states.\end{tabular} \\  & \\ \hline
\end{tabular}
\caption{Advantages and disadvantages of QR-PUFs compared to classical PUFs.}
\label{table2}
\end{table}

\section{Conclusion}
In this article, we proposed a theoretical framework for the quantitative characterisation of both PUFs and QR-PUFs. After developing an authentication protocol common to both typologies, with the same error correction and privacy amplification scheme, we formalized the $\text{(QR-)}$ PUFs in term of two main properties, the \emph{robustness} (connected to false rejection) and the \emph{unclonability} (connected to false acceptance). Finally, we studied some examples, motivating the possible advantages and disadvantages of QR-PUFs compared to classical PUFs. 

Our framework is useful to study and to compare different implementations of $\text{(QR-)}$ PUFs and to develop new authentication schemes.
An important application would be to strictly prove the superiority of QR-PUFs over classical PUFs. 
The next step towards that goal would be the development of new methods to estimate the unclonability of (QR-) PUFs for different implementations.
This could open an interesting line of theoretical and experimental research about $\text{(QR-)}$ PUFs.
Furthermore, our framework can be employed to determine the level of security of using $\text{(QR-)}$ PUFs in other cryptographic protocols, like QKD, where a quantitatively secure $\text{(QR-)}$ PUF can be used as authentication and reduces the number of necessary preshared key bits.

\emph{Note added:} During the finalisation of this work, we became aware of a preprint on a related topic \cite{ADDK}.

\section*{Acknowledgements}
The authors thank U. R{\u}hrmair for helpful discussions. This project has received funding from the German Federal Ministry of Education and Research (BMBF), within the Hardware-based Quantum Security (HQS) project.
\ifx\undefined\allcaps\def\allcaps#1{#1}\fi

\begin{thebibliography}{42}%
\makeatletter
\providecommand \@ifxundefined [1]{%
 \@ifx{#1\undefined}
}%
\providecommand \@ifnum [1]{%
 \ifnum #1\expandafter \@firstoftwo
 \else \expandafter \@secondoftwo
 \fi
}%
\providecommand \@ifx [1]{%
 \ifx #1\expandafter \@firstoftwo
 \else \expandafter \@secondoftwo
 \fi
}%
\providecommand \natexlab [1]{#1}%
\providecommand \enquote  [1]{``#1''}%
\providecommand \bibnamefont  [1]{#1}%
\providecommand \bibfnamefont [1]{#1}%
\providecommand \citenamefont [1]{#1}%
\providecommand \href@noop [0]{\@secondoftwo}%
\providecommand \href [0]{\begingroup \@sanitize@url \@href}%
\providecommand \@href[1]{\@@startlink{#1}\@@href}%
\providecommand \@@href[1]{\endgroup#1\@@endlink}%
\providecommand \@sanitize@url [0]{\catcode `\\12\catcode `\$12\catcode
  `\&12\catcode `\#12\catcode `\^12\catcode `\_12\catcode `\%12\relax}%
\providecommand \@@startlink[1]{}%
\providecommand \@@endlink[0]{}%
\providecommand \url  [0]{\begingroup\@sanitize@url \@url }%
\providecommand \@url [1]{\endgroup\@href {#1}{\urlprefix }}%
\providecommand \urlprefix  [0]{URL }%
\providecommand \Eprint [0]{\href }%
\providecommand \doibase [0]{https://doi.org/}%
\providecommand \selectlanguage [0]{\@gobble}%
\providecommand \bibinfo  [0]{\@secondoftwo}%
\providecommand \bibfield  [0]{\@secondoftwo}%
\providecommand \translation [1]{[#1]}%
\providecommand \BibitemOpen [0]{}%
\providecommand \bibitemStop [0]{}%
\providecommand \bibitemNoStop [0]{.\EOS\space}%
\providecommand \EOS [0]{\spacefactor3000\relax}%
\providecommand \BibitemShut  [1]{\csname bibitem#1\endcsname}%
\let\auto@bib@innerbib\@empty
\bibitem [{\citenamefont {Martin}(2012)}]{KM}%
  \BibitemOpen
  \bibfield  {author} {\bibinfo {author} {\bibfnamefont {K.~M.}\ \bibnamefont
  {Martin}},\ }\href@noop {} {\emph {\bibinfo {title} {Everyday Cryptography:
  Fundamental Principles and Applications}}}\ (\bibinfo  {publisher} {OUP
  Oxford},\ \bibinfo {year} {2012})\BibitemShut {NoStop}%
\bibitem [{\citenamefont {Scarani}\ \emph {et~al.}(2009)\citenamefont
  {Scarani}, \citenamefont {Bechmann-Pasquinucci}, \citenamefont {Cerf},
  \citenamefont {Du{\v{s}}ek}, \citenamefont {L{\"u}tkenhaus},\ and\
  \citenamefont {Peev}}]{SBCDLP}%
  \BibitemOpen
  \bibfield  {author} {\bibinfo {author} {\bibfnamefont {V.}~\bibnamefont
  {Scarani}}, \bibinfo {author} {\bibfnamefont {H.}~\bibnamefont
  {Bechmann-Pasquinucci}}, \bibinfo {author} {\bibfnamefont {N.~J.}\
  \bibnamefont {Cerf}}, \bibinfo {author} {\bibfnamefont {M.}~\bibnamefont
  {Du{\v{s}}ek}}, \bibinfo {author} {\bibfnamefont {N.}~\bibnamefont
  {L{\"u}tkenhaus}},\ and\ \bibinfo {author} {\bibfnamefont {M.}~\bibnamefont
  {Peev}},\ }\bibfield  {title} {\bibinfo {title} {The security of practical
  quantum key distribution},\ }\href@noop {} {\bibfield  {journal} {\bibinfo
  {journal} {Rev. Mod. Phys.}\ }\textbf {\bibinfo {volume} {81}},\ \bibinfo
  {pages} {1301} (\bibinfo {year} {2009})}\BibitemShut {NoStop}%
\bibitem [{\citenamefont {Wegman}\ and\ \citenamefont {Carter}(1981)}]{WC}%
  \BibitemOpen
  \bibfield  {author} {\bibinfo {author} {\bibfnamefont {M.~N.}\ \bibnamefont
  {Wegman}}\ and\ \bibinfo {author} {\bibfnamefont {J.~L.}\ \bibnamefont
  {Carter}},\ }\bibfield  {title} {\bibinfo {title} {New hash functions and
  their use in authentication and set equality},\ }\href@noop {} {\bibfield
  {journal} {\bibinfo  {journal} {J. Comput. Syst. Sci.}\ }\textbf {\bibinfo
  {volume} {22}},\ \bibinfo {pages} {265} (\bibinfo {year} {1981})}\BibitemShut
  {NoStop}%
\bibitem [{\citenamefont {Pappu}(2001)}]{RP}%
  \BibitemOpen
  \bibfield  {author} {\bibinfo {author} {\bibfnamefont {R.}~\bibnamefont
  {Pappu}},\ }\emph {\bibinfo {title} {Physical one-way functions}},\
  \href@noop {} {Ph.D. thesis},\ \bibinfo  {school} {Massachusetts Institute of
  Technology, USA} (\bibinfo {year} {2001})\BibitemShut {NoStop}%
\bibitem [{\citenamefont {R{\"u}hrmair}(2010)}]{UR10}%
  \BibitemOpen
  \bibfield  {author} {\bibinfo {author} {\bibfnamefont {U.}~\bibnamefont
  {R{\"u}hrmair}},\ }\bibfield  {title} {\bibinfo {title} {Oblivious transfer
  based on physical unclonable functions},\ }in\ \href@noop {} {\emph {\bibinfo
  {booktitle} {International Conference on Trust and Trustworthy Computing}}}\
  (\bibinfo {organization} {Springer},\ \bibinfo {year} {2010})\ pp.\ \bibinfo
  {pages} {430--440}\BibitemShut {NoStop}%
\bibitem [{\citenamefont {R{\"u}hrmair}\ and\ \citenamefont {van
  Dijk}(2013)}]{RD}%
  \BibitemOpen
  \bibfield  {author} {\bibinfo {author} {\bibfnamefont {U.}~\bibnamefont
  {R{\"u}hrmair}}\ and\ \bibinfo {author} {\bibfnamefont {M.}~\bibnamefont {van
  Dijk}},\ }\bibfield  {title} {\bibinfo {title} {On the practical use of
  physical unclonable functions in oblivious transfer and bit commitment
  protocols},\ }\href@noop {} {\bibfield  {journal} {\bibinfo  {journal} {J.
  Cryptogr. Eng.}\ }\textbf {\bibinfo {volume} {3}},\ \bibinfo {pages} {17}
  (\bibinfo {year} {2013})}\BibitemShut {NoStop}%
\bibitem [{\citenamefont {Brzuska}\ \emph {et~al.}(2011)\citenamefont
  {Brzuska}, \citenamefont {Fischlin}, \citenamefont {Schr{\"o}der},\ and\
  \citenamefont {Katzenbeisser}}]{BFSK}%
  \BibitemOpen
  \bibfield  {author} {\bibinfo {author} {\bibfnamefont {C.}~\bibnamefont
  {Brzuska}}, \bibinfo {author} {\bibfnamefont {M.}~\bibnamefont {Fischlin}},
  \bibinfo {author} {\bibfnamefont {H.}~\bibnamefont {Schr{\"o}der}},\ and\
  \bibinfo {author} {\bibfnamefont {S.}~\bibnamefont {Katzenbeisser}},\
  }\bibfield  {title} {\bibinfo {title} {Physically uncloneable functions in
  the universal composition framework},\ }in\ \href@noop {} {\emph {\bibinfo
  {booktitle} {Annual Cryptology Conference}}}\ (\bibinfo {organization}
  {Springer},\ \bibinfo {year} {2011})\ pp.\ \bibinfo {pages}
  {51--70}\BibitemShut {NoStop}%
\bibitem [{\citenamefont {Pappu}\ \emph {et~al.}(2002)\citenamefont {Pappu},
  \citenamefont {Recht}, \citenamefont {Taylor},\ and\ \citenamefont
  {Gershenfeld}}]{PRTG}%
  \BibitemOpen
  \bibfield  {author} {\bibinfo {author} {\bibfnamefont {R.}~\bibnamefont
  {Pappu}}, \bibinfo {author} {\bibfnamefont {B.}~\bibnamefont {Recht}},
  \bibinfo {author} {\bibfnamefont {J.}~\bibnamefont {Taylor}},\ and\ \bibinfo
  {author} {\bibfnamefont {N.}~\bibnamefont {Gershenfeld}},\ }\bibfield
  {title} {\bibinfo {title} {Physical one-way functions},\ }\href@noop {}
  {\bibfield  {journal} {\bibinfo  {journal} {Science}\ }\textbf {\bibinfo
  {volume} {297}},\ \bibinfo {pages} {2026} (\bibinfo {year}
  {2002})}\BibitemShut {NoStop}%
\bibitem [{\citenamefont {Lee}\ \emph {et~al.}(2004)\citenamefont {Lee},
  \citenamefont {Lim}, \citenamefont {Gassend}, \citenamefont {Suh},
  \citenamefont {Van~Dijk},\ and\ \citenamefont {Devadas}}]{LLGSDD}%
  \BibitemOpen
  \bibfield  {author} {\bibinfo {author} {\bibfnamefont {J.~W.}\ \bibnamefont
  {Lee}}, \bibinfo {author} {\bibfnamefont {D.}~\bibnamefont {Lim}}, \bibinfo
  {author} {\bibfnamefont {B.}~\bibnamefont {Gassend}}, \bibinfo {author}
  {\bibfnamefont {G.~E.}\ \bibnamefont {Suh}}, \bibinfo {author} {\bibfnamefont
  {M.}~\bibnamefont {Van~Dijk}},\ and\ \bibinfo {author} {\bibfnamefont
  {S.}~\bibnamefont {Devadas}},\ }\bibfield  {title} {\bibinfo {title} {A
  technique to build a secret key in integrated circuits for identification and
  authentication applications},\ }in\ \href@noop {} {\emph {\bibinfo
  {booktitle} {2004 Symposium on VLSI Circuits. Digest of Technical Papers
  (IEEE Cat. No. 04CH37525)}}}\ (\bibinfo {organization} {IEEE},\ \bibinfo
  {year} {2004})\ pp.\ \bibinfo {pages} {176--179}\BibitemShut {NoStop}%
\bibitem [{\citenamefont {Guajardo}\ \emph {et~al.}(2007)\citenamefont
  {Guajardo}, \citenamefont {Kumar}, \citenamefont {Schrijen},\ and\
  \citenamefont {Tuyls}}]{GKST}%
  \BibitemOpen
  \bibfield  {author} {\bibinfo {author} {\bibfnamefont {J.}~\bibnamefont
  {Guajardo}}, \bibinfo {author} {\bibfnamefont {S.~S.}\ \bibnamefont {Kumar}},
  \bibinfo {author} {\bibfnamefont {G.-J.}\ \bibnamefont {Schrijen}},\ and\
  \bibinfo {author} {\bibfnamefont {P.}~\bibnamefont {Tuyls}},\ }\bibfield
  {title} {\bibinfo {title} {$\text{FPGA}$ intrinsic $\text{PUFs}$ and their
  use for $\text{IP}$ protection},\ }in\ \href@noop {} {\emph {\bibinfo
  {booktitle} {International Workshop on Cryptographic Hardware and Embedded
  Systems}}}\ (\bibinfo {organization} {Springer},\ \bibinfo {year} {2007})\
  pp.\ \bibinfo {pages} {63--80}\BibitemShut {NoStop}%
\bibitem [{\citenamefont {Tuyls}\ \emph {et~al.}(2006)\citenamefont {Tuyls},
  \citenamefont {Schrijen}, \citenamefont {{\v{S}}kori{\'c}}, \citenamefont
  {Van~Geloven}, \citenamefont {Verhaegh},\ and\ \citenamefont
  {Wolters}}]{TSSGVW}%
  \BibitemOpen
  \bibfield  {author} {\bibinfo {author} {\bibfnamefont {P.}~\bibnamefont
  {Tuyls}}, \bibinfo {author} {\bibfnamefont {G.-J.}\ \bibnamefont {Schrijen}},
  \bibinfo {author} {\bibfnamefont {B.}~\bibnamefont {{\v{S}}kori{\'c}}},
  \bibinfo {author} {\bibfnamefont {J.}~\bibnamefont {Van~Geloven}}, \bibinfo
  {author} {\bibfnamefont {N.}~\bibnamefont {Verhaegh}},\ and\ \bibinfo
  {author} {\bibfnamefont {R.}~\bibnamefont {Wolters}},\ }\bibfield  {title}
  {\bibinfo {title} {Read-proof hardware from protective coatings},\ }in\
  \href@noop {} {\emph {\bibinfo {booktitle} {International Workshop on
  Cryptographic Hardware and Embedded Systems}}}\ (\bibinfo {organization}
  {Springer},\ \bibinfo {year} {2006})\ pp.\ \bibinfo {pages}
  {369--383}\BibitemShut {NoStop}%
\bibitem [{\citenamefont {Indeck}\ and\ \citenamefont {Muller}(1994)}]{IM}%
  \BibitemOpen
  \bibfield  {author} {\bibinfo {author} {\bibfnamefont {R.~S.}\ \bibnamefont
  {Indeck}}\ and\ \bibinfo {author} {\bibfnamefont {M.~W.}\ \bibnamefont
  {Muller}},\ }\href@noop {} {\bibinfo {title} {Method and apparatus for
  fingerprinting magnetic media}} (\bibinfo {year} {1994}),\ \bibinfo {note}
  {$\text{US}$ Patent 5, 365, 586}\BibitemShut {NoStop}%
\bibitem [{\citenamefont {Bossuet}\ \emph {et~al.}(2013)\citenamefont
  {Bossuet}, \citenamefont {Ngo}, \citenamefont {Cherif},\ and\ \citenamefont
  {Fischer}}]{BNCF}%
  \BibitemOpen
  \bibfield  {author} {\bibinfo {author} {\bibfnamefont {L.}~\bibnamefont
  {Bossuet}}, \bibinfo {author} {\bibfnamefont {X.~T.}\ \bibnamefont {Ngo}},
  \bibinfo {author} {\bibfnamefont {Z.}~\bibnamefont {Cherif}},\ and\ \bibinfo
  {author} {\bibfnamefont {V.}~\bibnamefont {Fischer}},\ }\bibfield  {title}
  {\bibinfo {title} {A $\text{PUF}$ based on a transient effect ring oscillator
  and insensitive to locking phenomenon},\ }\href@noop {} {\bibfield  {journal}
  {\bibinfo  {journal} {IEEE Trans. Emerg. Topics Comput.}\ }\textbf {\bibinfo
  {volume} {2}},\ \bibinfo {pages} {30} (\bibinfo {year} {2013})}\BibitemShut
  {NoStop}%
\bibitem [{\citenamefont {McGrath}\ \emph {et~al.}(2019)\citenamefont
  {McGrath}, \citenamefont {Bagci}, \citenamefont {Wang}, \citenamefont
  {Roedig},\ and\ \citenamefont {Young}}]{MBWRY}%
  \BibitemOpen
  \bibfield  {author} {\bibinfo {author} {\bibfnamefont {T.}~\bibnamefont
  {McGrath}}, \bibinfo {author} {\bibfnamefont {I.~E.}\ \bibnamefont {Bagci}},
  \bibinfo {author} {\bibfnamefont {Z.~M.}\ \bibnamefont {Wang}}, \bibinfo
  {author} {\bibfnamefont {U.}~\bibnamefont {Roedig}},\ and\ \bibinfo {author}
  {\bibfnamefont {R.~J.}\ \bibnamefont {Young}},\ }\bibfield  {title} {\bibinfo
  {title} {A $\text{PUF}$ taxonomy},\ }\href@noop {} {\bibfield  {journal}
  {\bibinfo  {journal} {Appl. Phys. Rev.}\ }\textbf {\bibinfo {volume} {6}},\
  \bibinfo {pages} {011303} (\bibinfo {year} {2019})}\BibitemShut {NoStop}%
\bibitem [{\citenamefont {Maes}\ and\ \citenamefont {Verbauwhede}(2010)}]{MV}%
  \BibitemOpen
  \bibfield  {author} {\bibinfo {author} {\bibfnamefont {R.}~\bibnamefont
  {Maes}}\ and\ \bibinfo {author} {\bibfnamefont {I.}~\bibnamefont
  {Verbauwhede}},\ }\bibfield  {title} {\bibinfo {title} {Physically unclonable
  functions: A study on the state of the art and future research directions},\
  }in\ \href@noop {} {\emph {\bibinfo {booktitle} {Towards Hardware-Intrinsic
  Security}}}\ (\bibinfo  {publisher} {Springer},\ \bibinfo {year} {2010})\
  pp.\ \bibinfo {pages} {3--37}\BibitemShut {NoStop}%
\bibitem [{\citenamefont {Delvaux}\ \emph {et~al.}(2014)\citenamefont
  {Delvaux}, \citenamefont {Gu}, \citenamefont {Schellekens},\ and\
  \citenamefont {Verbauwhede}}]{DGSV}%
  \BibitemOpen
  \bibfield  {author} {\bibinfo {author} {\bibfnamefont {J.}~\bibnamefont
  {Delvaux}}, \bibinfo {author} {\bibfnamefont {D.}~\bibnamefont {Gu}},
  \bibinfo {author} {\bibfnamefont {D.}~\bibnamefont {Schellekens}},\ and\
  \bibinfo {author} {\bibfnamefont {I.}~\bibnamefont {Verbauwhede}},\
  }\bibfield  {title} {\bibinfo {title} {Helper data algorithms for
  $\text{PUF}$-based key generation: Overview and analysis},\ }\href@noop {}
  {\bibfield  {journal} {\bibinfo  {journal} {IEEE Trans. Comput.-Aided Design
  Integr. Circuits Syst.}\ }\textbf {\bibinfo {volume} {34}},\ \bibinfo {pages}
  {889} (\bibinfo {year} {2014})}\BibitemShut {NoStop}%
\bibitem [{\citenamefont {Puchinger}\ \emph {et~al.}(2015)\citenamefont
  {Puchinger}, \citenamefont {M{\"u}elich}, \citenamefont {Bossert},
  \citenamefont {Hiller},\ and\ \citenamefont {Sigl}}]{PMBHS}%
  \BibitemOpen
  \bibfield  {author} {\bibinfo {author} {\bibfnamefont {S.}~\bibnamefont
  {Puchinger}}, \bibinfo {author} {\bibfnamefont {S.}~\bibnamefont
  {M{\"u}elich}}, \bibinfo {author} {\bibfnamefont {M.}~\bibnamefont
  {Bossert}}, \bibinfo {author} {\bibfnamefont {M.}~\bibnamefont {Hiller}},\
  and\ \bibinfo {author} {\bibfnamefont {G.}~\bibnamefont {Sigl}},\ }\bibfield
  {title} {\bibinfo {title} {On error correction for physical unclonable
  functions},\ }in\ \href@noop {} {\emph {\bibinfo {booktitle} {SCC 2015; 10th
  International ITG Conference on Systems, Communications and Coding}}}\
  (\bibinfo {organization} {VDE},\ \bibinfo {year} {2015})\ pp.\ \bibinfo
  {pages} {1--6}\BibitemShut {NoStop}%
\bibitem [{\citenamefont {Dodis}\ \emph {et~al.}(2008)\citenamefont {Dodis},
  \citenamefont {Ostrovsky}, \citenamefont {Reyzin},\ and\ \citenamefont
  {Smith}}]{DORS}%
  \BibitemOpen
  \bibfield  {author} {\bibinfo {author} {\bibfnamefont {Y.}~\bibnamefont
  {Dodis}}, \bibinfo {author} {\bibfnamefont {R.}~\bibnamefont {Ostrovsky}},
  \bibinfo {author} {\bibfnamefont {L.}~\bibnamefont {Reyzin}},\ and\ \bibinfo
  {author} {\bibfnamefont {A.}~\bibnamefont {Smith}},\ }\bibfield  {title}
  {\bibinfo {title} {Fuzzy extractors: How to generate strong keys from
  biometrics and other noisy data},\ }\href@noop {} {\bibfield  {journal}
  {\bibinfo  {journal} {SIAM J. Comput.}\ }\textbf {\bibinfo {volume} {38}},\
  \bibinfo {pages} {97} (\bibinfo {year} {2008})}\BibitemShut {NoStop}%
\bibitem [{\citenamefont {Helfmeier}\ \emph {et~al.}(2013)\citenamefont
  {Helfmeier}, \citenamefont {Boit}, \citenamefont {Nedospasov},\ and\
  \citenamefont {Seifert}}]{HBNS}%
  \BibitemOpen
  \bibfield  {author} {\bibinfo {author} {\bibfnamefont {C.}~\bibnamefont
  {Helfmeier}}, \bibinfo {author} {\bibfnamefont {C.}~\bibnamefont {Boit}},
  \bibinfo {author} {\bibfnamefont {D.}~\bibnamefont {Nedospasov}},\ and\
  \bibinfo {author} {\bibfnamefont {J.-P.}\ \bibnamefont {Seifert}},\
  }\bibfield  {title} {\bibinfo {title} {Cloning physically unclonable
  functions},\ }in\ \href@noop {} {\emph {\bibinfo {booktitle} {2013 IEEE
  International Symposium on Hardware-Oriented Security and Trust (HOST)}}}\
  (\bibinfo {organization} {IEEE},\ \bibinfo {year} {2013})\ pp.\ \bibinfo
  {pages} {1--6}\BibitemShut {NoStop}%
\bibitem [{\citenamefont {R{\"u}hrmair}\ \emph
  {et~al.}(2010{\natexlab{a}})\citenamefont {R{\"u}hrmair}, \citenamefont
  {Sehnke}, \citenamefont {S{\"o}lter}, \citenamefont {Dror}, \citenamefont
  {Devadas},\ and\ \citenamefont {Schmidhuber}}]{RSSDDS}%
  \BibitemOpen
  \bibfield  {author} {\bibinfo {author} {\bibfnamefont {U.}~\bibnamefont
  {R{\"u}hrmair}}, \bibinfo {author} {\bibfnamefont {F.}~\bibnamefont
  {Sehnke}}, \bibinfo {author} {\bibfnamefont {J.}~\bibnamefont {S{\"o}lter}},
  \bibinfo {author} {\bibfnamefont {G.}~\bibnamefont {Dror}}, \bibinfo {author}
  {\bibfnamefont {S.}~\bibnamefont {Devadas}},\ and\ \bibinfo {author}
  {\bibfnamefont {J.}~\bibnamefont {Schmidhuber}},\ }\bibfield  {title}
  {\bibinfo {title} {Modeling attacks on physical unclonable functions},\ }in\
  \href@noop {} {\emph {\bibinfo {booktitle} {Proceedings of the 17th ACM
  conference on Computer and communications security}}}\ (\bibinfo
  {organization} {ACM},\ \bibinfo {year} {2010})\ pp.\ \bibinfo {pages}
  {237--249}\BibitemShut {NoStop}%
\bibitem [{\citenamefont {R{\"u}hrmair}\ \emph {et~al.}(2013)\citenamefont
  {R{\"u}hrmair}, \citenamefont {S{\"o}lter}, \citenamefont {Sehnke},
  \citenamefont {Xu}, \citenamefont {Mahmoud}, \citenamefont {Stoyanova},
  \citenamefont {Dror}, \citenamefont {Schmidhuber}, \citenamefont {Burleson},\
  and\ \citenamefont {Devadas}}]{R-etal}%
  \BibitemOpen
  \bibfield  {author} {\bibinfo {author} {\bibfnamefont {U.}~\bibnamefont
  {R{\"u}hrmair}}, \bibinfo {author} {\bibfnamefont {J.}~\bibnamefont
  {S{\"o}lter}}, \bibinfo {author} {\bibfnamefont {F.}~\bibnamefont {Sehnke}},
  \bibinfo {author} {\bibfnamefont {X.}~\bibnamefont {Xu}}, \bibinfo {author}
  {\bibfnamefont {A.}~\bibnamefont {Mahmoud}}, \bibinfo {author} {\bibfnamefont
  {V.}~\bibnamefont {Stoyanova}}, \bibinfo {author} {\bibfnamefont
  {G.}~\bibnamefont {Dror}}, \bibinfo {author} {\bibfnamefont {J.}~\bibnamefont
  {Schmidhuber}}, \bibinfo {author} {\bibfnamefont {W.}~\bibnamefont
  {Burleson}},\ and\ \bibinfo {author} {\bibfnamefont {S.}~\bibnamefont
  {Devadas}},\ }\bibfield  {title} {\bibinfo {title} {$\text{PUF}$ modeling
  attacks on simulated and silicon data},\ }\href@noop {} {\bibfield  {journal}
  {\bibinfo  {journal} {IEEE Trans. Inf. Forensics Security}\ }\textbf
  {\bibinfo {volume} {8}},\ \bibinfo {pages} {1876} (\bibinfo {year}
  {2013})}\BibitemShut {NoStop}%
\bibitem [{\citenamefont {{\v{S}}kori{\'c}}(2012)}]{BS}%
  \BibitemOpen
  \bibfield  {author} {\bibinfo {author} {\bibfnamefont {B.}~\bibnamefont
  {{\v{S}}kori{\'c}}},\ }\bibfield  {title} {\bibinfo {title} {Quantum readout
  of physical unclonable functions},\ }\href@noop {} {\bibfield  {journal}
  {\bibinfo  {journal} {Int. J. Quantum Inf.}\ }\textbf {\bibinfo {volume}
  {10}},\ \bibinfo {pages} {1250001} (\bibinfo {year} {2012})}\BibitemShut
  {NoStop}%
\bibitem [{\citenamefont {Wootters}\ and\ \citenamefont {Zurek}(1982)}]{WZ}%
  \BibitemOpen
  \bibfield  {author} {\bibinfo {author} {\bibfnamefont {W.~K.}\ \bibnamefont
  {Wootters}}\ and\ \bibinfo {author} {\bibfnamefont {W.~H.}\ \bibnamefont
  {Zurek}},\ }\bibfield  {title} {\bibinfo {title} {A single quantum cannot be
  cloned},\ }\href@noop {} {\bibfield  {journal} {\bibinfo  {journal} {Nature}\
  }\textbf {\bibinfo {volume} {299}},\ \bibinfo {pages} {802} (\bibinfo {year}
  {1982})}\BibitemShut {NoStop}%
\bibitem [{\citenamefont {R{\"u}hrmair}\ \emph {et~al.}(2009)\citenamefont
  {R{\"u}hrmair}, \citenamefont {S{\"o}lter},\ and\ \citenamefont
  {Sehnke}}]{RSS}%
  \BibitemOpen
  \bibfield  {author} {\bibinfo {author} {\bibfnamefont {U.}~\bibnamefont
  {R{\"u}hrmair}}, \bibinfo {author} {\bibfnamefont {J.}~\bibnamefont
  {S{\"o}lter}},\ and\ \bibinfo {author} {\bibfnamefont {F.}~\bibnamefont
  {Sehnke}},\ }\bibfield  {title} {\bibinfo {title} {On the foundations of
  physical unclonable functions},\ }\href@noop {} {\bibfield  {journal}
  {\bibinfo  {journal} {IACR Cryptology ePrint Archive}\ }\textbf {\bibinfo
  {volume} {2009}},\ \bibinfo {pages} {277} (\bibinfo {year}
  {2009})}\BibitemShut {NoStop}%
\bibitem [{\citenamefont {Armknecht}\ \emph {et~al.}(2011)\citenamefont
  {Armknecht}, \citenamefont {Maes}, \citenamefont {Sadeghi}, \citenamefont
  {Standaert},\ and\ \citenamefont {Wachsmann}}]{AMSSW}%
  \BibitemOpen
  \bibfield  {author} {\bibinfo {author} {\bibfnamefont {F.}~\bibnamefont
  {Armknecht}}, \bibinfo {author} {\bibfnamefont {R.}~\bibnamefont {Maes}},
  \bibinfo {author} {\bibfnamefont {A.-R.}\ \bibnamefont {Sadeghi}}, \bibinfo
  {author} {\bibfnamefont {F.-X.}\ \bibnamefont {Standaert}},\ and\ \bibinfo
  {author} {\bibfnamefont {C.}~\bibnamefont {Wachsmann}},\ }\bibfield  {title}
  {\bibinfo {title} {A formalization of the security features of physical
  functions},\ }in\ \href@noop {} {\emph {\bibinfo {booktitle} {2011 IEEE
  Symposium on Security and Privacy}}}\ (\bibinfo {organization} {IEEE},\
  \bibinfo {year} {2011})\ pp.\ \bibinfo {pages} {397--412}\BibitemShut
  {NoStop}%
\bibitem [{\citenamefont {Plaga}\ and\ \citenamefont {Koob}(2012)}]{PK}%
  \BibitemOpen
  \bibfield  {author} {\bibinfo {author} {\bibfnamefont {R.}~\bibnamefont
  {Plaga}}\ and\ \bibinfo {author} {\bibfnamefont {F.}~\bibnamefont {Koob}},\
  }\bibfield  {title} {\bibinfo {title} {A formal definition and a new security
  mechanism of physical unclonable functions},\ }in\ \href@noop {} {\emph
  {\bibinfo {booktitle} {International GI/ITG Conference on Measurement,
  Modelling, and Evaluation of Computing Systems and Dependability and Fault
  Tolerance}}}\ (\bibinfo {organization} {Springer},\ \bibinfo {year} {2012})\
  pp.\ \bibinfo {pages} {288--301}\BibitemShut {NoStop}%
\bibitem [{\citenamefont {Plaga}\ and\ \citenamefont {Merli}(2015)}]{PM}%
  \BibitemOpen
  \bibfield  {author} {\bibinfo {author} {\bibfnamefont {R.}~\bibnamefont
  {Plaga}}\ and\ \bibinfo {author} {\bibfnamefont {D.}~\bibnamefont {Merli}},\
  }\bibfield  {title} {\bibinfo {title} {A new definition and classification of
  physical unclonable functions},\ }in\ \href@noop {} {\emph {\bibinfo
  {booktitle} {Proceedings of the Second Workshop on Cryptography and Security
  in Computing Systems}}}\ (\bibinfo {organization} {ACM},\ \bibinfo {year}
  {2015})\ p.~\bibinfo {pages} {7}\BibitemShut {NoStop}%
\bibitem [{\citenamefont {Delvaux}(2017)}]{JD}%
  \BibitemOpen
  \bibfield  {author} {\bibinfo {author} {\bibfnamefont {J.}~\bibnamefont
  {Delvaux}},\ }\emph {\bibinfo {title} {Security analysis of
  $\text{PUF}$-based key generation and entity authentication}},\ \href@noop {}
  {Ph.D. thesis},\ \bibinfo  {school} {Katholieke Universiteit Leuven, Belgium}
  (\bibinfo {year} {2017})\BibitemShut {NoStop}%
\bibitem [{\citenamefont {{\v{S}}kori{\'c}}\ \emph {et~al.}(2005)\citenamefont
  {{\v{S}}kori{\'c}}, \citenamefont {Tuyls},\ and\ \citenamefont
  {Ophey}}]{STO}%
  \BibitemOpen
  \bibfield  {author} {\bibinfo {author} {\bibfnamefont {B.}~\bibnamefont
  {{\v{S}}kori{\'c}}}, \bibinfo {author} {\bibfnamefont {P.}~\bibnamefont
  {Tuyls}},\ and\ \bibinfo {author} {\bibfnamefont {W.}~\bibnamefont {Ophey}},\
  }\bibfield  {title} {\bibinfo {title} {Robust key extraction from physical
  uncloneable functions},\ }in\ \href@noop {} {\emph {\bibinfo {booktitle}
  {International Conference on Applied Cryptography and Network Security}}}\
  (\bibinfo {organization} {Springer},\ \bibinfo {year} {2005})\ pp.\ \bibinfo
  {pages} {407--422}\BibitemShut {NoStop}%
\bibitem [{\citenamefont {Goorden}\ \emph {et~al.}(2014)\citenamefont
  {Goorden}, \citenamefont {Horstmann}, \citenamefont {Mosk}, \citenamefont
  {{\v{S}}kori{\'c}},\ and\ \citenamefont {Pinkse}}]{GHMSP}%
  \BibitemOpen
  \bibfield  {author} {\bibinfo {author} {\bibfnamefont {S.~A.}\ \bibnamefont
  {Goorden}}, \bibinfo {author} {\bibfnamefont {M.}~\bibnamefont {Horstmann}},
  \bibinfo {author} {\bibfnamefont {A.~P.}\ \bibnamefont {Mosk}}, \bibinfo
  {author} {\bibfnamefont {B.}~\bibnamefont {{\v{S}}kori{\'c}}},\ and\ \bibinfo
  {author} {\bibfnamefont {P.~W.}\ \bibnamefont {Pinkse}},\ }\bibfield  {title}
  {\bibinfo {title} {Quantum-secure authentication of a physical unclonable
  key},\ }\href@noop {} {\bibfield  {journal} {\bibinfo  {journal} {Optica}\
  }\textbf {\bibinfo {volume} {1}},\ \bibinfo {pages} {421} (\bibinfo {year}
  {2014})}\BibitemShut {NoStop}%
\bibitem [{\citenamefont {Tuyls}\ \emph {et~al.}(2005)\citenamefont {Tuyls},
  \citenamefont {{\v{S}}kori{\'c}}, \citenamefont {Stallinga}, \citenamefont
  {Akkermans},\ and\ \citenamefont {Ophey}}]{TSSAO}%
  \BibitemOpen
  \bibfield  {author} {\bibinfo {author} {\bibfnamefont {P.}~\bibnamefont
  {Tuyls}}, \bibinfo {author} {\bibfnamefont {B.}~\bibnamefont
  {{\v{S}}kori{\'c}}}, \bibinfo {author} {\bibfnamefont {S.}~\bibnamefont
  {Stallinga}}, \bibinfo {author} {\bibfnamefont {A.~H.}\ \bibnamefont
  {Akkermans}},\ and\ \bibinfo {author} {\bibfnamefont {W.}~\bibnamefont
  {Ophey}},\ }\bibfield  {title} {\bibinfo {title} {Information-theoretic
  security analysis of physical uncloneable functions},\ }in\ \href@noop {}
  {\emph {\bibinfo {booktitle} {International Conference on Financial
  Cryptography and Data Security}}}\ (\bibinfo {organization} {Springer},\
  \bibinfo {year} {2005})\ pp.\ \bibinfo {pages} {141--155}\BibitemShut
  {NoStop}%
\bibitem [{\citenamefont {Rioul}\ \emph {et~al.}(2016)\citenamefont {Rioul},
  \citenamefont {Sol{\'e}}, \citenamefont {Guilley},\ and\ \citenamefont
  {Danger}}]{RSGD}%
  \BibitemOpen
  \bibfield  {author} {\bibinfo {author} {\bibfnamefont {O.}~\bibnamefont
  {Rioul}}, \bibinfo {author} {\bibfnamefont {P.}~\bibnamefont {Sol{\'e}}},
  \bibinfo {author} {\bibfnamefont {S.}~\bibnamefont {Guilley}},\ and\ \bibinfo
  {author} {\bibfnamefont {J.-L.}\ \bibnamefont {Danger}},\ }\bibfield  {title}
  {\bibinfo {title} {On the entropy of physically unclonable functions},\ }in\
  \href@noop {} {\emph {\bibinfo {booktitle} {2016 IEEE International Symposium
  on Information Theory (ISIT)}}}\ (\bibinfo {organization} {IEEE},\ \bibinfo
  {year} {2016})\ pp.\ \bibinfo {pages} {2928--2932}\BibitemShut {NoStop}%
\bibitem [{\citenamefont {{\v{S}}kori{\'c}}\ \emph {et~al.}(2013)\citenamefont
  {{\v{S}}kori{\'c}}, \citenamefont {Mosk},\ and\ \citenamefont
  {Pinkse}}]{SMP}%
  \BibitemOpen
  \bibfield  {author} {\bibinfo {author} {\bibfnamefont {B.}~\bibnamefont
  {{\v{S}}kori{\'c}}}, \bibinfo {author} {\bibfnamefont {A.~P.}\ \bibnamefont
  {Mosk}},\ and\ \bibinfo {author} {\bibfnamefont {P.~W.}\ \bibnamefont
  {Pinkse}},\ }\bibfield  {title} {\bibinfo {title} {Security of
  quantum-readout $\text{PUFs}$ against quadrature-based challenge-estimation
  attacks},\ }\href@noop {} {\bibfield  {journal} {\bibinfo  {journal} {Int. J.
  Quantum Inf.}\ }\textbf {\bibinfo {volume} {11}},\ \bibinfo {pages} {1350041}
  (\bibinfo {year} {2013})}\BibitemShut {NoStop}%
\bibitem [{\citenamefont {{\v{S}}koric}(2016)}]{BS13}%
  \BibitemOpen
  \bibfield  {author} {\bibinfo {author} {\bibfnamefont {B.}~\bibnamefont
  {{\v{S}}koric}},\ }\bibfield  {title} {\bibinfo {title} {Security analysis of
  quantum-readout $\text{PUFs}$ in the case of challenge-estimation attacks},\
  }\href@noop {} {\bibfield  {journal} {\bibinfo  {journal} {Quantum Inf.
  Comput.}\ }\textbf {\bibinfo {volume} {16}},\ \bibinfo {pages} {0050}
  (\bibinfo {year} {2016})}\BibitemShut {NoStop}%
\bibitem [{\citenamefont {Yao}\ \emph {et~al.}(2016)\citenamefont {Yao},
  \citenamefont {Gao}, \citenamefont {Li},\ and\ \citenamefont {Zhang}}]{YGLZ}%
  \BibitemOpen
  \bibfield  {author} {\bibinfo {author} {\bibfnamefont {Y.}~\bibnamefont
  {Yao}}, \bibinfo {author} {\bibfnamefont {M.}~\bibnamefont {Gao}}, \bibinfo
  {author} {\bibfnamefont {M.}~\bibnamefont {Li}},\ and\ \bibinfo {author}
  {\bibfnamefont {J.}~\bibnamefont {Zhang}},\ }\bibfield  {title} {\bibinfo
  {title} {Quantum cloning attacks against $\text{PUF}$-based quantum
  authentication systems},\ }\href@noop {} {\bibfield  {journal} {\bibinfo
  {journal} {Quantum Inf. Process.}\ }\textbf {\bibinfo {volume} {15}},\
  \bibinfo {pages} {3311} (\bibinfo {year} {2016})}\BibitemShut {NoStop}%
\bibitem [{\citenamefont {Nikolopoulos}\ and\ \citenamefont
  {Diamanti}(2017)}]{ND}%
  \BibitemOpen
  \bibfield  {author} {\bibinfo {author} {\bibfnamefont {G.~M.}\ \bibnamefont
  {Nikolopoulos}}\ and\ \bibinfo {author} {\bibfnamefont {E.}~\bibnamefont
  {Diamanti}},\ }\bibfield  {title} {\bibinfo {title} {Continuous-variable
  quantum authentication of physical unclonable keys},\ }\href@noop {}
  {\bibfield  {journal} {\bibinfo  {journal} {Sci. Rep.}\ }\textbf {\bibinfo
  {volume} {7}},\ \bibinfo {pages} {46047} (\bibinfo {year}
  {2017})}\BibitemShut {NoStop}%
\bibitem [{\citenamefont {Nikolopoulos}(2018)}]{GN}%
  \BibitemOpen
  \bibfield  {author} {\bibinfo {author} {\bibfnamefont {G.~M.}\ \bibnamefont
  {Nikolopoulos}},\ }\bibfield  {title} {\bibinfo {title} {Continuous-variable
  quantum authentication of physical unclonable keys: Security against an
  emulation attack},\ }\href@noop {} {\bibfield  {journal} {\bibinfo  {journal}
  {Phys. Rev. A}\ }\textbf {\bibinfo {volume} {97}},\ \bibinfo {pages} {012324}
  (\bibinfo {year} {2018})}\BibitemShut {NoStop}%
\bibitem [{\citenamefont {Zyczkowski}\ and\ \citenamefont {Kus}(1994)}]{ZK}%
  \BibitemOpen
  \bibfield  {author} {\bibinfo {author} {\bibfnamefont {K.}~\bibnamefont
  {Zyczkowski}}\ and\ \bibinfo {author} {\bibfnamefont {M.}~\bibnamefont
  {Kus}},\ }\bibfield  {title} {\bibinfo {title} {Random unitary matrices},\
  }\href@noop {} {\bibfield  {journal} {\bibinfo  {journal} {J. Phys. A: Math.
  Gen.}\ }\textbf {\bibinfo {volume} {27}},\ \bibinfo {pages} {4235} (\bibinfo
  {year} {1994})}\BibitemShut {NoStop}%
\bibitem [{\citenamefont {Gross}\ \emph {et~al.}(2010)\citenamefont {Gross},
  \citenamefont {Liu}, \citenamefont {Flammia}, \citenamefont {Becker},\ and\
  \citenamefont {Eisert}}]{GLFBE}%
  \BibitemOpen
  \bibfield  {author} {\bibinfo {author} {\bibfnamefont {D.}~\bibnamefont
  {Gross}}, \bibinfo {author} {\bibfnamefont {Y.-K.}\ \bibnamefont {Liu}},
  \bibinfo {author} {\bibfnamefont {S.~T.}\ \bibnamefont {Flammia}}, \bibinfo
  {author} {\bibfnamefont {S.}~\bibnamefont {Becker}},\ and\ \bibinfo {author}
  {\bibfnamefont {J.}~\bibnamefont {Eisert}},\ }\bibfield  {title} {\bibinfo
  {title} {Quantum state tomography via compressed sensing},\ }\href@noop {}
  {\bibfield  {journal} {\bibinfo  {journal} {Phys. Rev. Lett.}\ }\textbf
  {\bibinfo {volume} {105}},\ \bibinfo {pages} {150401} (\bibinfo {year}
  {2010})}\BibitemShut {NoStop}%
\bibitem [{\citenamefont {R{\"u}hrmair}\ \emph
  {et~al.}(2010{\natexlab{b}})\citenamefont {R{\"u}hrmair}, \citenamefont
  {Busch},\ and\ \citenamefont {Katzenbeisser}}]{RBK}%
  \BibitemOpen
  \bibfield  {author} {\bibinfo {author} {\bibfnamefont {U.}~\bibnamefont
  {R{\"u}hrmair}}, \bibinfo {author} {\bibfnamefont {H.}~\bibnamefont
  {Busch}},\ and\ \bibinfo {author} {\bibfnamefont {S.}~\bibnamefont
  {Katzenbeisser}},\ }\bibfield  {title} {\bibinfo {title} {Strong
  $\text{PUFs}$: models, constructions, and security proofs},\ }in\ \href@noop
  {} {\emph {\bibinfo {booktitle} {Towards Hardware-Intrinsic Security}}}\
  (\bibinfo  {publisher} {Springer},\ \bibinfo {year} {2010})\ pp.\ \bibinfo
  {pages} {79--96}\BibitemShut {NoStop}%
\bibitem [{\citenamefont {Canetti}\ \emph {et~al.}(2016)\citenamefont
  {Canetti}, \citenamefont {Fuller}, \citenamefont {Paneth}, \citenamefont
  {Reyzin},\ and\ \citenamefont {Smith}}]{CFPRS}%
  \BibitemOpen
  \bibfield  {author} {\bibinfo {author} {\bibfnamefont {R.}~\bibnamefont
  {Canetti}}, \bibinfo {author} {\bibfnamefont {B.}~\bibnamefont {Fuller}},
  \bibinfo {author} {\bibfnamefont {O.}~\bibnamefont {Paneth}}, \bibinfo
  {author} {\bibfnamefont {L.}~\bibnamefont {Reyzin}},\ and\ \bibinfo {author}
  {\bibfnamefont {A.}~\bibnamefont {Smith}},\ }\bibfield  {title} {\bibinfo
  {title} {Reusable fuzzy extractors for low-entropy distributions},\ }in\
  \href@noop {} {\emph {\bibinfo {booktitle} {Annual International Conference
  on the Theory and Applications of Cryptographic Techniques}}}\ (\bibinfo
  {organization} {Springer},\ \bibinfo {year} {2016})\ pp.\ \bibinfo {pages}
  {117--146}\BibitemShut {NoStop}%
\bibitem [{\citenamefont {Helstrom}(1969)}]{CH}%
  \BibitemOpen
  \bibfield  {author} {\bibinfo {author} {\bibfnamefont {C. W.}~\bibnamefont
  {Helstrom}},\ }\bibfield  {title} {\bibinfo {title} {Quantum detection and estimation theory},\ }\href@noop {} {\bibfield  {journal}
  {\bibinfo  {journal} {J. Stat. Phys.}\ }\textbf {\bibinfo {volume}
  {1}},\ \bibinfo {pages} {231--252} (\bibinfo {year} {1969})}\BibitemShut
  {NoStop}%
\bibitem [{\citenamefont {Bru{\ss}}\ and\ \citenamefont
  {Macchiavello}(2001)}]{BM01}%
  \BibitemOpen
  \bibfield  {author} {\bibinfo {author} {\bibfnamefont {D.}~\bibnamefont
  {Bru{\ss}}}\ and\ \bibinfo {author} {\bibfnamefont {C.}~\bibnamefont
  {Macchiavello}},\ }\bibfield  {title} {\bibinfo {title} {Optimal cloning for
  two pairs of orthogonal states},\ }\href@noop {} {\bibfield  {journal}
  {\bibinfo  {journal} {J. Phys. A: Math. Gen.}\ }\textbf {\bibinfo {volume}
  {34}},\ \bibinfo {pages} {6815} (\bibinfo {year} {2001})}\BibitemShut
  {NoStop}%
\bibitem{ADDK}
M. Arapinis, M. Delavar, M. Doosti and E. Kashefi,
Quantum Physical Unclonable Functions: Possibilities and Impossibilities,
arXiv:1910.02126v1 (2019).
\end{thebibliography}
\end{document}